\documentclass{article}

\usepackage[usenames,dvipsnames]{xcolor}
\usepackage{amsmath,amssymb,tikz,xcolor,dsfont,enumerate,subcaption}

\usepackage{setspace,url}

\newcommand{\ds}{\displaystyle}

\begin{document}

\title{Causal Effects in Twin Studies: the Role of Interference }

\author{Bonnie Smith$^1$, Elizabeth L. Ogburn$^1$, Matt McGue$^2$, \\ Saonli Basu$^3$, Daniel O. Scharfstein$^1$}

\date{$^1$ Department of Biostatistics, Johns Hopkins Bloomberg School of Public Health, Baltimore, MD \\
$^2$ Department of Psychology, University of Minnesota, Minneapolis, MN \\
$^3$ Division of Biostatistics, University of Minnesota, Minneapolis, MN  \\
bsmit179@jhmi.edu}

\maketitle

\abstract{The use of twins designs to address causal questions is becoming increasingly popular.  A standard assumption is that there is no interference between twins---that is, no twin's exposure has a causal impact on their co-twin's outcome.  However, there may be settings in which this assumption would not hold, and this would (1) impact the causal interpretation of parameters obtained by commonly used existing methods;  (2) change which effects are of greatest interest;  and (3) impact the conditions under which we may estimate these effects.  We explore these issues, and we derive semi-parametric efficient estimators for causal effects in the presence of interference between twins.  Using data from the Minnesota Twin Family Study, we apply our estimators to assess whether twins' consumption of alcohol in early adolescence may have a causal impact on their co-twins' substance use later in life.}

\bigskip

\noindent \textbf{Keywords:}  Co-twin control method; Semi-parametric efficiency;  Spillover effect

%%%%%%%%%%%%%%%%%%%%%%

\section{Introduction}

While researchers have predominantly used twin studies to assess the heritability of a given trait,  they are also increasingly leveraging twin designs to learn about the causal relationship between an exposure and an outcome:  see for example \cite{McGue_utility, Irons,Johnson,Kendler,Lahey,Schaefer} and references therein.   A popular tool in this area of twins research is the co-twin control method, in which an unexposed twin serves as the control for their exposed co-twin.  This method uses the fact that twins are naturally matched on many predictors of the exposure and the outcome, such as shared genetics and characteristics of their shared environment;  this can make it possible to estimate the causal effect of the exposure on the outcome even if these shared predictors are unobserved \cite{McGue_utility}.  In this paper, we draw attention to an important causal assumption that underlies this approach.  This is the assumption that there is no \emph{interference} between twins, meaning one twin's exposure has no causal impact on their co-twin's outcome.  This is a standard assumption made in the twins literature \cite{Sjolander,Lahey};  however, evidence from the literature on sibling influence suggests that it may not hold in some cases.  A number of studies have identified variables, such as substance use, where a subject's behavior may be causally impacted by that of their siblings \cite{Slomkowski,Lauersen,Oettinger}.  Since there is evidence of interference between siblings in these settings, it seems tenable that between twins, especially, interference could be present in some cases.

The possibility of interference between twins is important for a number of reasons.  One is that the parameter estimated using the co-twin control method has a different causal interpretation when interference is present.   Incorrectly assuming that there is no interference may therefore lead one to draw misleading conclusions.  Interference impacts several issues regarding the causal parameter to be studied, including which parameters are of most scientific interest, and conditions under which these causal parameters can be identified from the observed data.  If interference is present, key causal effects of interest include \emph{spillover effects}, which are changes that would result in twins' outcomes from intervening on their co-twins' exposures, while holding their own exposures fixed;  and \emph{main effects}, which are changes that would result in twins' outcomes from intervening on their own exposures, while holding their co-twins' exposures fixed.   Unfortunately, the co-twin control method does not provide a means of estimating spillover effects or main effects when there is interference between twins;  rather, it targets an effect of more limited scientific and policy relevance.

In this paper, we clarify which causal effect is estimated by the co-twin control method when there is interference between twins, and argue that it is not a causal effect of prime interest.  In order to identify effects that are of prime of interest, such as main effects and spillover effects, we then proceed under an assumption that the measured baseline covariates control for all confounding of the effects of the exposures on the outcomes.  In this framework, we derive estimators of average main effects and average spillover effects which are \emph{semi-parametric efficient}---that is, they make the most efficient use of the data possible, given the assumptions of the statistical models that we use.   

In Section 2, we define potential outcomes and causal effects for our setting of independent pairs of twins with possible between-twin interference, and we review the co-twin control method for the case where there is no interference.  In Section 3, we consider the co-twin control method with interference, then introduce the causal models that we will use in the rest of the paper, which make the assumption of no unmeasured confounding.  In Section 4 we present the semi-parametric efficient estimators of average spillover effects and average main effects in these models.  In Section 5 we apply these estimators to data from the Minnesota Twin Family Study, and investigate whether twins' exposure to alcohol in early adolescence may have a causal impact on their co-twins' drinking behavior in adulthood.  We demonstrate the finite-sample performance of our estimators in a simulation study in Section 6, and conclude with a discussion in Section 7.

%%%%%%%%%%%%%%%%%%%%%%%%%%%%%

\section{Background}

\subsection{Causal effects for the setting of within-pairs interference}

Consider a twin study with a binary exposure $A$, where $A=1$ if the subject is exposed and $A=0$ if the subject is unexposed, and an outcome $Y$, and suppose first that there is no interference between subjects.  In this setting each subject is considered to have two \emph{potential outcomes}:  $Y^1$, the outcome that the subject would have if, possibly counter to fact, they were to have the exposure;  and $Y^0$, the outcome that the subject would have if, possibly counter to fact, they were to be unexposed.   The observed outcome $Y$ is assumed to be equal to the potential outcome $Y^1$ for subjects who have exposure $A=1$, and equal to $Y^0$ for subjects with $A=0$.  While only one potential outcome is observed for each twin---and hence the causal effect $Y^1-Y^0$ for an individual is never known---under additional assumptions (of positivity and exchangeability), the population average of these effects can be estimated from the observed data. Common targets of inference in twin studies include the average causal effect $E \big[ Y^1 - Y^0 \big]$, where the mean is over all twins in the target population, and the average causal effect in the subgroup of twins who are discordant with their co-twin for the exposure.

Now suppose instead that there may be interference between the two twins in each pair, but not between twins from different twin pairs, so that each twin's outcome may be impacted by their own exposure and their co-twin's exposure.  In this case, we cannot meaningfully talk about a twin's outcome if they were to be exposed, or $Y^1$, since they might have one outcome if both they and their co-twin were to be exposed, and a different outcome if they were to be exposed but their co-twin were not.  Thus, in this setting, each twin has 4 potential outcomes:  $Y^{1,1}$, $Y^{1,0}$, $Y^{0,1}$, and $Y^{0,0}$, where we write $Y^{a,b}$ for the outcome that the twin would have if, possibly counter to fact, they were to have exposure $a$ and their co-twin were to have exposure $b$.  For a twin who has exposure $A=a$ and whose co-twin has exposure $A=b$, the observed outcome $Y$ is assumed to be equal to the potential outcome $Y^{a,b}$, while the twin's other 3 potential outcomes are not observed.  There are a number of different causal effects that we can consider in this setting:  of particular interest are \emph{average main effects} $E \big[ Y^{0,0}-Y^{1,0} \big]$ and $E \big[ Y^{0,1}-Y^{1,1} \big]$, in which the subject's own exposure is varied while their co-twin's exposure is held constant;  and  \emph{average spillover effects} $E \big[ Y^{0,0}-Y^{0,1} \big]$ and $E \big[ Y^{1,0}-Y^{1,1} \big]$, in which the subject's own exposure is held constant while their co-twin's exposure is varied.  

Methods for estimating average main effects and average spillover effects have been studied by several authors.   One body of research considers the setting of \emph{partial interference}, in which there are multiple distinct groups of subjects and interference does not operate across different groups.  See for example \cite{HongRaudenbush, HudgensHalloran, TchetgenVanderWeele, LiuIPW, LiuDR} for estimation of average main effects and average spillover effects under partial interference, and see \cite{LiuHudgens,LiuIPW,LiuDR} for asymptotic properties of the estimators.  A different strand of the interference literature considers the setting where there is only one group, such as a single connected social network where interference could occur between any of the subjects.  See for example \cite{Aronow,VanDerLaan, Ogburn, Miles, AutoG}.  The context that we consider here, namely independent groups of size two, falls into the first of these two frameworks.  To the best of our knowledge, no previous work on partial interference demonstrated semi-parametric efficiency of an estimator.  Here we derive semi-parametric efficient estimators for two models that are tailored to the setting of independent pairs.

%%%%%%%%%%%%%%%%%%%%%%%%%%%%%%%%%

\subsection{Notation and data structure}

In describing the data for a twin pair, we distinguish between those baseline covariates which are characteristics of the pair---and thus necessarily common between the two twins in a given pair---such as zygosity or parental characteristics; and baseline covariates which are characteristics of an individual twin and may or may not be the same for the two twins in a given pair.  We refer to these respectively as shared covariates and individual covariates.  In each twin pair, suppose that the twins have been randomly labeled as Twin 1 and Twin 2.   The observed data for a given pair is $O=\big( C,X_1,X_2,A_1,A_2,Y_1,Y_2)$, where $C$ denotes the shared baseline covariates, $X_j$ denotes the individual baseline covariates for Twin $j$, and $A_j$ and $Y_j$ are the binary exposure and the outcome for Twin $j$.   Throughout, $C$, $X_1$, and $X_2$ will refer to collections of measured baseline covariates.  In some sections we will also consider factors which are unobserved;   we will use notation such as $U$ or $U_{sh}$ when we refer to collections of variables at least some of which are unobserved.  We assume that we observe data for $n$ twin pairs, and that the observed data $O_1,\ldots,O_n$ for these $n$ pairs are independent and identically distributed.

We use the notation $Y_j^{a,b}$ for a potential outcome for Twin $j$:  specifically, let $Y_j^{a,b}$ be the outcome that Twin $j$ would have if, possibly counter to fact, Twin $j$ were to have exposure $a$ and their co-twin were to have exposure $b$.  Our primary targets of inference will be means of potential outcomes $E \big[ Y^{a,b} \big]$ and contrasts of such parameters, where here the mean is taken over all twins in the population.  We can also write the $E \big[ Y^{a,b} \big]$ as $E \big[ \frac{1}{2} \big(Y_1^{a,b} + Y_2^{a,b}\big) \big]$, where $\frac{1}{2} \big(Y_1^{a,b} + Y_2^{a,b}\big)$ is the average of the potential outcomes within a twin pair, and the last expectation is a mean taken over all pairs.   We will often express $E \big[ Y^{a,b} \big]$ this way, so that the units we work in are the independent twin pairs rather than the individual twins.

%%%%%%%%%%%%%%%%%%%%%%%%%%%%%%%%%%%

\subsection{The co-twin control method under the assumption of no interference}

Here we review the co-twin control method for the setting where there is no interference between twins, following Sj\"{o}lander et al. \cite{Sjolander} and McGue et al. \cite{McGue_utility};  in Section 3 we compare the setting where there is interference.

The causal effect of the exposure $A_j$ on the outcome $Y_j$ may be confounded by any predictor $V$ that is a common cause of both $A_j$ and $Y_j$ (since there will be a non-causal source of association between $A_j$ and $Y_j$ via the path through $V$).  Such a predictor could be a measured or unmeasured factor individual to Twin $j$, or a measured or unmeasured factor shared by both twins.  While we can explicitly adjust for measured confounders, unobserved confounders would typically preclude identification of the causal effect.  However, using the co-twin control method, all shared factors---whether measured or unmeasured---are naturally accounted for by the fact that the two twins are matched on these factors.  Therefore, Sj\"{o}lander et al.\cite{Sjolander} showed, in cases where there is no unmeasured confounding due to individual factors, a causal effect is identified by adjusting for the measured individual covariates $X_1$ and $X_2$.  

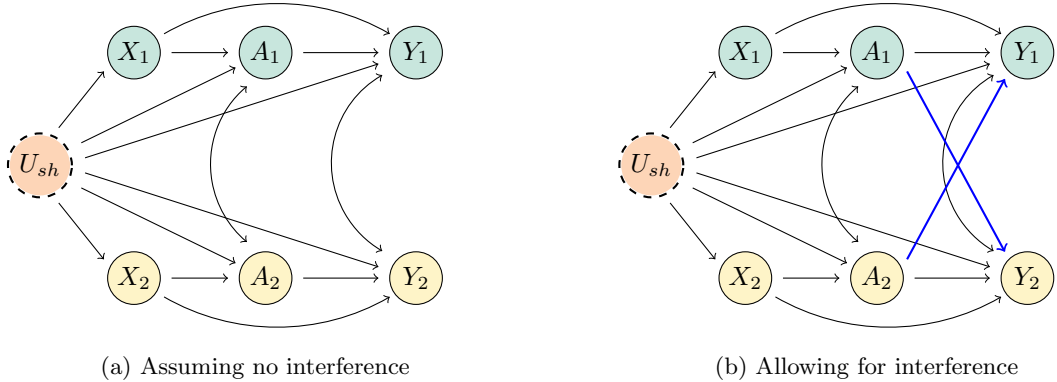
\begin{figure}

\begin{center}
\begin{subfigure}{.4\textwidth}

\begin{tikzpicture}[scale=.5]

\fill[Orange!30!white] (-3,0) circle (.8);
\draw[thick,dashed] (-3,0) circle (.85);
\draw (-3,0) node{$U_{sh}$};

\draw[->] (-2.5,1)--(-1.3,2.5);
\draw[->] (-2.5,-1)--(-1.3,-2.5);

\fill[SeaGreen!30!white] (-.5,3) circle (.7);
\draw (-.5,3) circle (.7);
\draw (-.5,3) node{$X_1$};

\draw[->] (.5,3)--(2,3);

\fill[Goldenrod!30!white] (-.5,-3) circle (.7);
\draw (-.5,-3) circle (.7);
\draw (-.5,-3) node{$X_2$};

\draw[->] (.5,-3)--(2,-3);

\draw[->] (-1.8,.2)--(6,2.7);
\draw[->] (-1.8,-.2)--(6,-2.7);

\draw[->] (-1.9,.6)--(2.2,2.6);
\draw[->] (-1.9,-.6)--(2.2,-2.6);

\fill[SeaGreen!30!white] (3,3) circle (.7);
\draw (3,3) circle (.7);
\draw (3,3) node{$A_1$};

\fill[Goldenrod!30!white] (3,-3) circle (.7);
\draw (3,-3) circle (.7);
\draw (3,-3) node{$A_2$};

\draw[<->] (2.5,2.1) arc(130:230:2.7);

\fill[SeaGreen!30!white] (7,3) circle (.7);
\draw (7,3) circle (.7);
\draw (7,3) node{$Y_1$};

\draw[->] (4,3)--(6,3);
\draw[->] (4,-3)--(6,-3);

\fill[Goldenrod!30!white] (7,-3) circle (.7);
\draw (7,-3) circle (.7);
\draw (7,-3) node{$Y_2$};

\draw[<->] (6.1,2.4) arc(120:240:2.7);

\draw[->] (.3,3.5) arc(120:60:6);
\draw[->] (.3,-3.5) arc(240:300:6);

\end{tikzpicture}

\caption{Assuming no interference}
\label{no_interference_ctc}

\end{subfigure}
%%%%%%%%%%%%%%%%
\qquad \qquad
\begin{subfigure}{.4\textwidth}

\begin{tikzpicture}[scale=.5]

\fill[Orange!30!white] (-3,0) circle (.8);
\draw[thick,dashed] (-3,0) circle (.85);
\draw (-3,0) node{$U_{sh}$};

\draw[->] (-2.5,1)--(-1.3,2.5);
\draw[->] (-2.5,-1)--(-1.3,-2.5);

\fill[SeaGreen!30!white] (-.5,3) circle (.7);
\draw (-.5,3) circle (.7);
\draw (-.5,3) node{$X_1$};

\draw[->] (.5,3)--(2,3);

\fill[Goldenrod!30!white] (-.5,-3) circle (.7);
\draw (-.5,-3) circle (.7);
\draw (-.5,-3) node{$X_2$};

\draw[->] (.5,-3)--(2,-3);

\draw[->] (-1.8,.2)--(6,2.7);
\draw[->] (-1.8,-.2)--(6,-2.7);

\draw[->] (-1.9,.6)--(2.2,2.6);
\draw[->] (-1.9,-.6)--(2.2,-2.6);

\fill[SeaGreen!30!white] (3,3) circle (.7);
\draw (3,3) circle (.7);
\draw (3,3) node{$A_1$};

\fill[Goldenrod!30!white] (3,-3) circle (.7);
\draw (3,-3) circle (.7);
\draw (3,-3) node{$A_2$};

\draw[<->] (2.5,2.1) arc(130:230:2.7);

\fill[SeaGreen!30!white] (7,3) circle (.7);
\draw (7,3) circle (.7);
\draw (7,3) node{$Y_1$};

\draw[->] (4,3)--(6,3);
\draw[->] (4,-3)--(6,-3);

\fill[Goldenrod!30!white] (7,-3) circle (.7);
\draw (7,-3) circle (.7);
\draw (7,-3) node{$Y_2$};

\draw[<->] (6.1,2.4) arc(120:240:2.7);

\draw[thick,blue, ->]  (3.8,2.5)--(6.4,-2.3);
\draw[ thick,blue,->]  (3.8,-2.5)--(6.4,2.3);

\draw[->] (.3,3.5) arc(120:60:6);
\draw[->] (.3,-3.5) arc(240:300:6);

\end{tikzpicture}

\caption{Allowing for interference}
\label{interference_ctc}

\end{subfigure}

\caption{Causal diagrams for two settings where the co-twin control method can be used.  Here $U_{sh}$ represents all confounders, both measured and unmeasured, which are shared by the two twins, while $X_j$ are measured individual covariates for Twin $j$, and $A_j$ and $Y_j$ are Twin $j$'s exposure and outcome.  The blue arrows in (b) allow for possible interference between the two twins, i.e. one twin's exposure may have a causal impact on their co-twin's outcome.}

\end{center}
\end{figure}

Let $U_{sh}$ denote the set of all confounders, including both measured and unmeasured factors, which are shared between the two twins.  Consider the causal diagram in Figure \ref{no_interference_ctc}, where a directed arrow means that the first variable has a possible causal impact on the second, while a bidirected arc between two variables indicates that there may be unobserved variables that are common causes of the two variables.  Assume that the relationship among the variables is given by this diagram.  In particular, assume that there is no interference between the two twins, as signaled by the absence of directed arrows $A_1 \to Y_2$ and $A_2 \to Y_1$, and assume that there are no non-shared unmeasured variables that directly impact both $A_j$ and $Y_j$.  

Consider the subgroup of twins who are discordant with their co-twins for the exposure, but who have the same level $x$ of individual covariates as their co-twins.  Write $E \big[ Y^1-Y^0 | A_1 \neq A_2, X_1=X_2=x \big]$ to denote the average causal effect on this subgroup.  For an exposed twin in this subgroup, their potential outcome $Y^1$ is observed; and while their potential outcome $Y^0$ is not observed, their unexposed co-twin's potential outcome $Y^0$ is observed.   \cite{Sjolander} show that, because of symmetry between the group of twins designated Twin 1 and the group designated Twin 2, we may use the unexposed co-twins' outcomes as proxies for the exposed twins' $Y^0$ potential outcomes, and that the causal parameter $E \big[ Y^1-Y^0 | A_1 \neq A_2, X_1=X_2=x \big]$ is equal to $E \big[ Y_j | A_j=1, A_{3-j}=0, X_1=X_2=x \big] - E \big[ Y_j | A_j=0, A_{3-j}=1, X_1=X_2=x \big]$, the mean difference in the observed outcome for the exposed twin and the observed outcome for the unexposed twin, within this subgroup.  The latter difference can now be estimated, for example by fitting a between-within regression model as described in Sj\"{o}lander et al.   Thus when all individual confounders are fully observed, Sj\"{o}lander et al. have shown that the within-pair coefficient in the between-within regression model has a causal interpretation, which is the causal effect for the subgroup of twins who are discordant with their co-twin for the exposure but have the same level of individual covariates as their co-twin.

%%%%%%%%%%%%%%%%%%%%%%%%%  Figure
%%%%%%%%%%%%%%%%%%%%%%%%%%%%%%%%%%%%

%%%%%%%%%%%%%%%%%%%%%%%%%%%%%%%%%%%%%%%%%%%%%%

\section{Identification of causal effects when interference is present}

Now we consider a setting where interference between twins may exist, that is, where it is possible for one twin's exposure to affect their co-twin's outcome.  For example, this could occur because the two twins influence one another's behavior or share information related to the exposure with each other.  Interference is likely to be present in many studies where the exposure and outcome are behavioral;  and incorrectly assuming that there is no interference can lead to misleading conclusions, and ignores spillover effects which may often be of interest.

%%%%%%%%%%%%%%%%%%%%%

\subsection{The co-twin control effect when there is interference between twins}

Here we modify the scenario described in Section 2.3 to allow for interference between the two twins in each pair.  Suppose now that the relationship among the variables is as in Figure \ref{interference_ctc}.  In particular, there may be shared factors (both measured and unmeasured) which impact both the exposures and the outcomes, but we assume that there are no unmeasured non-shared factors that directly cause both $A_j$ and $Y_j$, or that directly cause both $A_{3-j}$ and $Y_j$.  As before, the matching between the twins on shared factors allows for identification of one subgroup causal effect:  importantly, this is the effect which is estimated using the co-twin control method if there is interference present.  However, we argue that this is not an effect of prime importance, and that a different approach is needed in order to identify more useful effects such as average main effects and average spillover effects.

%%%%%%%%%%%%%%%%%%%%%%%%%% Table

\renewcommand{\arraystretch}{2}

\begin{table}
\begin{center}
\begin{tabular}{|c|c|}
\hline
Between-within regression model & \hspace{-.5in} $E\big[ Y_j | A_j,A_{3-j},X_j,X_{3-j} \big]= $ \\
\  & \hspace{.25in} $\beta_0 + \beta_W A_j + \beta_B \overline{A} + \gamma X_j + \delta X_{3-j}$ \\
\hline
Statistical value of the $\beta_W$ & \hspace{-.25in} $E \big[ Y_j | A_j=1, A_{3-j}=0, X_1=X_2 \big] - $ \\
regression coefficient & \hspace{0.25in} $E \big[ Y_j | A_j=0, A_{3-j}=1, X_1=X_2 \big]$ \\
\hline
Causal interpretation of $\beta_W$ with & $E \big[ Y^1-Y^0 | A_1 \neq A_2, X_1=X_2 \big]$ \\
no interference between twins  & \ \\
\hline
Causal interpretation of $\beta_W$ with  &  $E \big[ Y^{1,0}-Y^{0,1} | A_1 \neq A_2, X_1=X_2 \big]$ \\
interference between twins & \  \\
\hline
\end{tabular}
\caption{Comparison of the causal interpretation of the within-pair coefficient $\beta_W$ in a between-within regression model, for the settings with and without interference between twins.  In the no-interference case, $Y^1-Y^0$ is the difference in a twin's outcome under an intervention changing the twin from unexposed to exposed.  In the interference case, $Y^{1,0}-Y^{0,1}$ is the difference in a twin's outcome under an intervention changing the twin from unexposed to exposed, while conversely changing their co-twin from exposed to unexposed.}
\label{ctc_interpretation}
\end{center}
\end{table}
%%%%%%%%%%%%%%%%%%%%%%%%%%%%%%%%

As in Section 2.3, consider the subgroup of twins who are discordant with their co-twins for the exposure, but who have the same level $x$ of individual covariates as their co-twins.  For an exposed twin in this subgroup, their potential outcome $Y^{1,0}$ is observed, and for an unexposed twin in this subgroup, their potential outcome $Y^{0,1}$ is observed.  In Appendix A, we prove that, under the assumptions listed there, the mean of the contrast $Y^{1,0}-Y^{0,1}$ on this subgroup, $E \big[ Y^{1,0}-Y^{0,1} | A_1 \neq A_2, X_1=X_2=x \big]$, is equal to $E \big[ Y_j | A_j=1, A_{3-j}=0, X_1=X_2=x \big] - E \big[ Y_j | A_j=0, A_{3-j}=1, X_1=X_2=x \big]$.    Note that the latter difference is the same quantity that appears in Section 2.3, and which is estimated using a between-within regression model, but that the causal interpretation of this parameter is different in cases where there is interference between twins.  The different interpretations for the two settings are highlighted in Table \ref{ctc_interpretation}.  The effect $E \big[ Y^{1,0}-Y^{0,1} | A_1 \neq A_2, X_1=X_2=x \big]$ identified here is the average difference in twins' outcomes that we would see if we could intervene to change the twins from unexposed to exposed, while conversely intervening to change their co-twins from exposed to unexposed, in the subgroup described above.  It is difficult to see why one would want to target this particular combination of interventions.  The contrast $Y^{1,0}-Y^{0,1}$ is the difference of the spillover effect $Y^{0,0} - Y^{0,1}$ and the main effect $ Y^{0,0} - Y^{1,0}$;  however we cannot tease apart these two effects, and the value of $Y^{1,0}-Y^{0,1}$ itself is not readily interpretable.  For example, a zero value of $Y^{1,0}-Y^{0,1}$ could reflect the fact that $Y^{1,0}=Y^{0,0}=Y^{0,1}$, or it could reflect a qualitatively different scenario where $Y^{0,0}-Y^{0,1}$ and $Y^{0,0}-Y^{1,0}$ are each nonzero but cancel each other out.  That is, a null value of $Y^{1,0}-Y^{0,1}$ is equally compatible with the exposure having no causal effect, or with the presence of very strong interference.

The co-twin control method is a unique tool which allows for estimation of a causal effect even with shared unmeasured confounders, based on the use of discordant pairs.  However, we cannot identify contrasts involving $Y^{1,1}$ or $Y^{0,0}$ from discordant pairs when interference is present, since these potential outcomes need not equal the observed outcome of either twin in a discordant pair.  Therefore, the presence of shared unmeasured confounders poses a greater barrier than it does in settings with no interference, since the co-twin control method does not provide a means of estimating important causal effects for the interference setting.  In order to identify average main effects and average spillover effects, throughout the rest of the paper we work under the assumption that there is no unmeasured confounding due to either individual or shared factors.

%%%%%%%%%%%%%%%%%%%%%%%%%%%%%%%%%%%

\subsection{Identification under the assumption of no unmeasured confounding}

Throughout the rest of the paper we make the following assumption of \emph{no unmeasured confounding}:  there are no unmeasured factors, whether shared or non-shared, which directly cause both $A_j$ and $Y_j$, or that directly cause both $A_{3-j}$ and $Y_j$.  Specifically, we assume that any unmeasured factors $U$ which impact both an outcome and an exposure do so only through the measured baseline covariates $C$, $X_1$, and $X_2$.  Under this assumption, adjusting for the measured baseline covariates $C,X_1,X_2$ controls for all confounding of the effects of $A_j$ and $A_{3-j}$ on $Y_j$.  This is an untestable assumption which is not automatically satisfied by design in an observational study;  how reasonable the assumption is in a given study will depend on factors specific to that study, including how rich a set of baseline covariates is measured.

Consider the four groups of twin pairs with the four exposure patterns $(A_1=1,A_2=1)$, $(A_1=1,A_2=0)$, $(A_1=0,A_2=1)$, and $(A_1=0,A_2=0)$, and consider a specific potential outcome $Y_j^{a,b}$.   In general the potential outcomes $Y_j^{a,b}$ could be systematically higher among one of these four groups than another.  For example they could be higher among the $(A_1=1,A_2=1)$ group than among the $(A_1=0,A_2=0)$ group if having some predictor $V$ causes both exposures and potential outcomes to be higher.  However, our assumption of no unmeasured confounding implies that adjusting for the measured baseline covariates breaks any such association between $Y_j^{a,b}$ and the exposures, and that within levels of the measured baseline covariates $C,X_1,X_2$, the four groups are exchangeable in terms of potential outcomes $Y_j^{a,b}$.  This assumption of no unmeasured confounding, or \emph{exchangeability}, is given by:
\begin{align*}
 \mbox{\textbf{A1}}:  & \ \ Y_j^{a,b} \perp \!\!\! \perp (A_1,A_2 ) \ \big| \ ( C, X_1,X_2 ) \ \mbox{ for each fixed } a,b=0,1, \ j=1,2 & \hspace{0.25in} \mbox{ \emph{(exchangeability) }}
 \end{align*}
We additionally make the positivity assumption that, within each level of the baseline covariates, there is a nonzero probability of having each of the 4 exposure patterns, and the consistency assumption that a twin's observed outcome $Y_j$ is equal to the twin's potential outcome corresponding to the exposures actually received by the the twin and their co-twin:
\begin{align*}
\textbf{A2}: & \  \mbox{ For all } c,x_1,x_2 \mbox{ in the support of } C,X_1,X_2,  & \mbox{ \emph{(positivity)}} \\
\ & \ P \big(A_1=a,A_2=b \ | \ C=c,X_1=x_1,X_2=x_2 \big) > 0 \mbox{ for all } a,b=0,1. \\
\textbf{A3}: & \ \mbox{ If } A_j=a \mbox{ and } A_{3-j}=b, \mbox{ then } Y_j=Y_j^{a,b} & \mbox{\emph{(consistency)}}
\end{align*}
The assumptions of exchangeability, positivity, and consistency are sufficient for identification of the parameters $E \big[ Y_j^{a,b} \big]$.  That is, under these three assumptions, we can express the causal parameter $E \big[ Y_j^{a,b} \big]$ as a function of the observed data, as we show for completeness below.  Exchangeability and positivity allow us to take the mean over just the group with the exposure pattern $(A_j=a,A_{3-j}=b)$ rather than over all twins labeled as Twin $j$ in the population, within each level of the baseline covariates;  and consistency implies that, among the twins in this group, the potential outcome $Y_j^{a,b}$ is equal to the observed outcome $Y_j$. 
\begin{align*}
 E \big[  Y_j^{a,b}  \big]  =  & \ \int_y  y \ dF \big( Y_j^{a,b}=y \big) & \ \\
 = & \  \int_{c,x_1,x_2} \int_{y} y \ dF \big( Y_j^{a,b} = y \big| c,x_1,x_2 \big)  dF(c,x_1,x_2) & \ \\
 = & \  \int_{c,x_1,x_2}  \int_{y} y \  dF \big( Y_j^{a,b} =y \big| A_j=a,A_{3-j}=b,c,x_1,x_2 \big)  dF(c,x_1,x_2) &  \mbox{ by \textbf{A1} and \textbf{A2} } \\
 = & \  \int_{c,x_1,x_2}  \int_y y \ dF \big( Y_j =y \big| A_j=a,A_{3-j}=b,c,x_1,x_2 \big) dF(c,x_1,x_2)  &  \mbox{ by \textbf{A3} }  \\
 =& \ E \Big[ E \big[ Y_j \ \big| \ A_j=a,A_{3-j}=b, C,X_1,X_2 \big] \Big] .
 \end{align*}
 
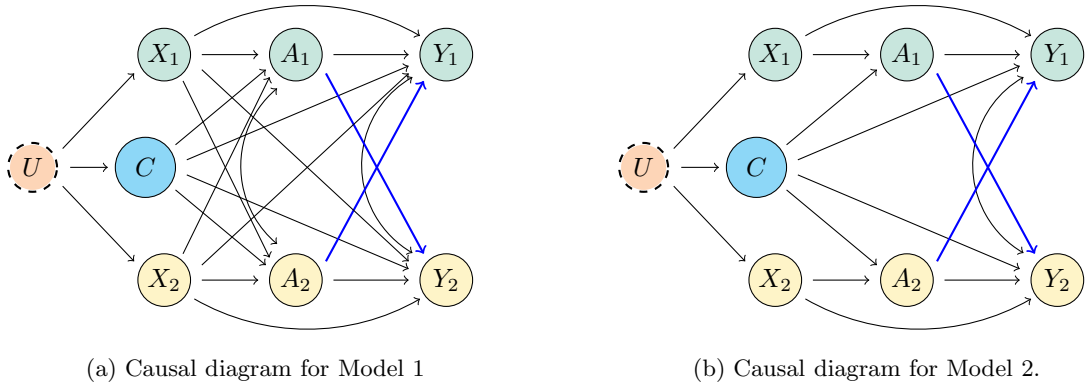
\begin{figure}
\begin{center}

\begin{subfigure}{.4\textwidth}

\begin{tikzpicture}[scale=.5]

\fill[Orange!30!white] (-4,0) circle (.6);
\draw[thick,dashed] (-4,0) circle (.65);
\draw (-4,0) node{$U$};

\draw[->] (-3,0)--(-2,0);
\draw[->] (-3.2,.5)--(-1.3,2.5);
\draw[->] (-3.2,-.5)--(-1.3,-2.5);

\fill[Cyan!40!white] (-1,0) circle (.8);
\draw (-1,0) circle (.8);
\draw (-1,0) node{$C$};

\fill[SeaGreen!30!white] (-.5,3) circle (.7);
\draw (-.5,3) circle (.7);
\draw (-.5,3) node{$X_1$};

\draw[->] (.5,3)--(2,3);

\fill[Goldenrod!30!white] (-.5,-3) circle (.7);
\draw (-.5,-3) circle (.7);
\draw (-.5,-3) node{$X_2$};

\draw[->] (.5,-3)--(2,-3);

\draw[->] (.1,.2)--(6,2.7);
\draw[->] (.1,-.2)--(6,-2.7);

\draw[->] (-.2,.6)--(2.2,2.6);
\draw[->] (-.2,-.6)--(2.2,-2.6);

\fill[SeaGreen!30!white] (3,3) circle (.7);
\draw (3,3) circle (.7);
\draw (3,3) node{$A_1$};

\fill[Goldenrod!30!white] (3,-3) circle (.7);
\draw (3,-3) circle (.7);
\draw (3,-3) node{$A_2$};

\draw[<->] (2.5,2.1) arc(130:230:2.7);

\fill[SeaGreen!30!white] (7,3) circle (.7);
\draw (7,3) circle (.7);
\draw (7,3) node{$Y_1$};

\draw[->] (4,3)--(6,3);
\draw[->] (4,-3)--(6,-3);

\fill[Goldenrod!30!white] (7,-3) circle (.7);
\draw (7,-3) circle (.7);
\draw (7,-3) node{$Y_2$};

\draw[<->] (6.1,2.4) arc(120:240:2.7);

\draw[thick,blue, ->]  (3.8,2.5)--(6.4,-2.3);
\draw[ thick,blue,->]  (3.8,-2.5)--(6.4,2.3);

\draw[->] (.5,2.6)--(6,-2.5);
\draw[->] (0,2.3)--(2.3,-2.4);

\draw[->] (.5,-2.6)--(6,2.5);
\draw[->] (0,-2.3)--(2.3,2.4);

\draw[->] (.3,3.5) arc(120:60:6);
\draw[->] (.3,-3.5) arc(240:300:6);

\end{tikzpicture}

\caption{Causal diagram for Model 1}
\label{Model1}

\end{subfigure}
%%%%%%%%%%%%%%%%%%%
\qquad \qquad 
\begin{subfigure}{.4\textwidth}

\begin{tikzpicture}[scale=.5]

\fill[Orange!30!white] (-4,0) circle (.6);
\draw[thick,dashed] (-4,0) circle (.65);
\draw (-4,0) node{$U$};

\draw[->] (-3,0)--(-2,0);
\draw[->] (-3.2,.5)--(-1.3,2.5);
\draw[->] (-3.2,-.5)--(-1.3,-2.5);

\fill[Cyan!40!white] (-1,0) circle (.8);
\draw (-1,0) circle (.8);
\draw (-1,0) node{$C$};

\fill[SeaGreen!30!white] (-.5,3) circle (.7);
\draw (-.5,3) circle (.7);
\draw (-.5,3) node{$X_1$};

\draw[->] (.5,3)--(2,3);

\fill[Goldenrod!30!white] (-.5,-3) circle (.7);
\draw (-.5,-3) circle (.7);
\draw (-.5,-3) node{$X_2$};

\draw[->] (.5,-3)--(2,-3);

\draw[->] (.1,.2)--(6,2.7);
\draw[->] (.1,-.2)--(6,-2.7);

\draw[->] (-.2,.6)--(2.2,2.6);
\draw[->] (-.2,-.6)--(2.2,-2.6);

\fill[SeaGreen!30!white] (3,3) circle (.7);
\draw (3,3) circle (.7);
\draw (3,3) node{$A_1$};

\fill[Goldenrod!30!white] (3,-3) circle (.7);
\draw (3,-3) circle (.7);
\draw (3,-3) node{$A_2$};

\fill[SeaGreen!30!white] (7,3) circle (.7);
\draw (7,3) circle (.7);
\draw (7,3) node{$Y_1$};

\draw[->] (4,3)--(6,3);
\draw[->] (4,-3)--(6,-3);

\fill[Goldenrod!30!white] (7,-3) circle (.7);
\draw (7,-3) circle (.7);
\draw (7,-3) node{$Y_2$};

\draw[<->] (6.1,2.4) arc(120:240:2.7);

\draw[thick,blue, ->]  (3.8,2.5)--(6.4,-2.3);
\draw[ thick,blue,->]  (3.8,-2.5)--(6.4,2.3);

\draw[->] (.3,3.5) arc(120:60:6);
\draw[->] (.3,-3.5) arc(240:300:6);

\end{tikzpicture}

\caption{Causal diagram for Model 2.}
\label{Model2}

\end{subfigure}

\caption{Causal diagrams for two models allowing between-twin interference but making the assumption of no unobserved confounding.  $C,X_1,X_2$ are measured baseline covariates, while here $U$ represents unmeasured factors (shared and/or non-shared).   In Model 2 we also assume that a twin's individual covariates do not have a causal impact on their co-twin's exposure or outcome, and that Twin 1's and Twin 2's exposures are conditionally independent given observed covariates.}

\end{center}
\end{figure}

 We will consider two models for the data for each twin pair.  The larger of these, Model 1, makes only the assumptions used for identification above, and corresponds to the diagram in Figure \ref{Model1}.   Here $U$ represents any unmeasured factors (shared or not).  The absence of arrows pointing directly from $U$ to the exposures and the outcomes corresponds to the no unmeasured confounding assumption.  Model 2 is a smaller model in which we make three extra assumptions in addition to the identification assumptions, and corresponds to the diagram in Figure \ref{Model2}.  These three extra assumptions, \textbf{A4}-\textbf{A6}, are that Twin $j$'s individual covariates do not have a causal impact on their co-twin's exposure (as seen by the lack of a directed arrow $X_j \to A_{3-j}$), or on their co-twin's outcome (as seen by the lack of a directed arrow $X_j \to Y_{3-j}$);  and that Twin 1's exposure and Twin 2's exposure are conditionally independent given the measured shared and individual baseline covariates (as seen by the lack of a bidirected arc $A_1 \leftrightarrow A_2$).  
\begin{align*}
\mbox{\textbf{A4}:} & \ A_j \perp \!\!\! \perp X_{3-j}  \ | \ (C,X_{j} \big) \mbox{ for } a, b = 0,1, \ j=1,2  \\
\mbox{\textbf{A5}:} &  \ Y_j^{a,b} \perp \!\!\! \perp X_{3-j}  \ | \ (C,X_{j} \big) \mbox{ for all } a, b = 0,1, j=1,2 & \mbox{\emph{(Model 2 assumptions)} } \\
\mbox{\textbf{A6}:}  & \ A_1 \perp \!\!\! \perp A_2 \ | \ (C,X_1,X_2 \big)
\end{align*}
The assumption that Twin $j$'s individual-level covariates do not impact their co-twin's exposure or outcome is one that is commonly used in the twin literature \cite{Sjolander}, and the distinction drawn in Model 2 between the individual and the shared covariates is a feature that distinguishes Model 2 from models considered elsewhere in the interference literature.  While Model 1, being less restrictive, gives valid inference in a wider range of settings than Model 2, the advantage of Model 2 is that it allows for improved (asymptotic) efficiency:   in settings which meet the assumptions posited for Model 2, we may leverage these additional assumptions to make a more efficient use of the data in estimating our causal parameters of interest, allowing for narrower confidence intervals in large samples.  We demonstrate this improved efficiency in a simulation study in Section 6.

%%%%%%%%%%%%%%%%%%%%%%%%%%%%%% Figure
%%%%%%%%%%%%%%%%%%%%%%%%%%%%%%%%%%%

 %%%%%%%%%%%%%%%%%%%%%%%%%%%%%%%%%%%%%

\section{Efficient estimation of causal effects when interference is present}

\subsection{Efficient estimators}

Here we derive the efficient estimator of the parameter $\beta^{a,b} =E \big[ Y^{a,b} \big]$ for each of the two models described in Section 3.2.   That is, we derive the estimator $\widehat{\beta}_1^{a,b}$ with the smallest asymptotic variance, out of all estimators for $\beta^{a,b}$ which are regular and asymptotically linear (RAL) under every distribution in Model 1, and similarly for Model 2.  Since Model 2 is a smaller model contained in Model 1, any RAL estimator for Model 1 may be used in Model 2, while the converse need not hold;  and we will see that, in fact, there are RAL estimators for Model 2 that are more efficient than any Model 1 estimator. 

RAL estimators are asymptotically normal with asymptotic variance determined by their influence function.   In order to obtain the RAL estimator with the smallest possible asymptotic variance, we first derive the \emph{efficient influence function} $\varphi_k( O ; \beta^{a,b})$ for the parameter $\beta^{a,b}$ in each Model $k$.  The efficient estimator $\widehat{\beta}_k^{a,b}$ based on data $O_1,\ldots,O_n$ from $n$ independent twin pairs is then found as the solution to the estimating equation $\sum_{i=1}^n \varphi_k(O_i ; \beta^{a,b} ) = 0$.  See Tsiatis \cite{Tsiatis} for more on efficient influence functions.   

The efficient influence function for $\beta^{a,b} = E \big[ Y^{a,b} \big]$ in Model  1 is:
 
 \begin{align*} \varphi_1 \big( O ; \beta^{a,b} \big)= & \  \frac{1}{2} \Bigg\{  \frac{ \mathds{1}(A_1=a,A_2=b)}{P(A_1=a,A_2=b | C,X_1,X_2) }\Big(Y_1- E \big[ Y_1 | A_1=a,A_2=b,C,X_1,X_2 \big] \Big)+ \\
& \hspace{.5in}  E \big[ Y_1 | A_1=a,A_2=b,C,X_1,X_2  \big]  \Bigg\} + \\
& \ \frac{1}{2} \Bigg\{ \frac{ \mathds{1}(A_2=a,A_1=b)}{P(A_2=a,A_1=b | C,X_1,X_2 ) }\Big(Y_2- E \big[ Y_2 | A_2=a,A_1=b,C,X_1,X_2  \big] \Big)+ \\
& \hspace{.5in}  E \big[ Y_2 | A_2=a,A_1=b,C,X_1,X_2  \big] \Bigg\} - \beta^{a,b} .
\end{align*}

Proofs are given in Appendix B.  In order to use $\varphi_1 \big( O ; \beta^{a,b} \big)$ to obtain an estimator for $\beta^{a,b}$, we need a model for the joint propensity score $P(A_j=a,A_{3-j}=b | C,X_j,X_{3-j})$ and a model for the outcome regression $E \big[ Y_j | A_j,A_{3-j}, C,X_j,X_{3-j} \big]$.  The resulting estimator of $\beta^{a,b}$ will be efficient provided that both of these models are correctly specified, and provided that the estimated values converge to the truth at fast enough rates.  (See Section 4.2 for more discussion of rates.)  Suppose this is the case.  Let $\hat{\pi}_{a,b}(C_i,X_{ij},X_{i,3-j})$ be the predicted propensity score for a twin pair with covariates $C_i,X_{ij},X_{i,3-j}$, and let $\hat{\mu}_{j,a,b}(C,X_{ij},X_{i,3-j})$ be the predicted outcome regression for Twin $j$ in a twin pair with covariates $C_i,X_{ij},X_{i,3-j}$ if the twins' exposures were set to $A_j=a,A_{3-j}=b$.  Then the efficient estimator for $\beta^{a,b}$ in Model 1, based on data $O_1,\ldots,O_n$, is given by:
 \begin{align*} 
 \widehat{\beta}^{a,b}_1 =  \ \frac{1}{n} \sum_{i=1}^n  \ \frac{1}{2} \Bigg\{ & \frac{ \mathds{1}(A_{i1}=a,A_{i2}=b)}{ \hat{\pi}_{a,b}( C_i,X_{i1},X_{i2} ) }\Big(Y_{i1}- \hat{\mu}_{1,a,b}(C_i,X_{i1},X_{i2} ) \Big)+ \hat{\mu}_{1,a,b}(C_i,X_{i1},X_{i2} ) \\
& + \ \frac{ \mathds{1}(A_{i1}=b,A_{i2}=a)}{ \hat{\pi}_{a,b}( C_i,X_{i2},X_{i1} ) }\Big(Y_{i2}- \hat{\mu}_{2,a,b}(C_i,X_{i2},X_{i1} ) \Big)+ \hat{\mu}_{2,a,b}(C_i,X_{i2},X_{i1} ) \Bigg\} .
\end{align*}

The estimator $\widehat{\beta}_1^{a,b}$ is the augmented inverse probability weighted estimator for a bivariate exposure.  It is a doubly robust estimator \cite{Scharfstein99}:  as long as at least one of the propensity score model or the outcome regression model is correctly specified, $\widehat{\beta}_1^{a,b}$ remains a consistent estimator for $\beta^{a,b}$, even if the other model is misspecified.  Liu et al. \cite{LiuDR} have presented three doubly robust estimators of average main effects and average spillover effects in groups of size $n$, working in the analog of Model 1.  Our estimator $\widehat{\beta}_1^{a,b}$ corresponds to one of their estimators specialized to groups of size 2.  Our contribution as regards Model 1 is to prove that this estimator is semiparametric efficient in Model 1, thus providing a partial answer to a question posed in Liu et al. \cite{LiuDR}.  To the best of our knowledge, Model 2 is distinct from other models that have been considered in the interference literature.  

If the true distribution lies inside the smaller Model 2, a more efficient estimator than $\widehat{\beta}_1^{a,b}$ is possible.  The efficient influence function for $\beta^{a,b}$ in Model  2 is:

\begin{align*} \varphi_2 \big( O ; \beta^{a,b} \big)=  & \  \frac{1}{2} \Bigg\{  \frac{ \mathds{1}(A_1=a,A_2=b)}{P(A_1=a | C,X_1) P(A_2=b|C,X_1)}\Big(Y_1- E \big[ Y_1 | A_1=a,A_2=b,C,X_1  \big] \Big)+ \\
& \hspace{.5in}  E \big[ Y_1 | A_1=a,A_2=b,C,X_1 \big]  \Bigg\} + \\
&  \ \frac{1}{2} \Bigg\{ \frac{ \mathds{1}(A_2=a,A_1=b)}{P(A_2=a | C,X_2)P(A_1=b|C,X_2 ) }\Big(Y_2- E \big[ Y_2 | A_2=a,A_1=b,C,X_2  \big] \Big)+ \\
& \hspace{.5in}  E \big[ Y_2 | A_2=a,A_1=b,C,X_2 \big] \Bigg\} - \beta^{a,b} .
\end{align*}

In order to use $\varphi_2(O ; \beta^{a,b})$ to obtain an estimator of $\beta^{a,b}$, we need a model for the outcome regression $E \big[ Y_j | A_j, A_{3-j}, C, X_j \big]$, a model for the propensity score $P(A_j =1 | C, X_j)$ which relates Twin $j$'s exposure to their own covariates, and a model for the propensity score $P(A_j=1 | C, X_{3-j})$ which relates Twin $j$'s exposure to their co-twin's covariates.  The resulting estimator of $\beta^{a,b}$ will be efficient provided that all three of these models are correctly specified, and provided that the estimated values converge to the truth at fast enough rates.  Suppose this is the case.  Let $\hat{\pi}_{j,a}(C_i,X_{ij})$ be the predicted value of the propensity score $P(A_j=a | C,X_j)$ for a twin pair with covariates $C_i,X_{ij}$.  Let $\hat{\theta}_{j,a}(C_i,X_{i,3-j})$ be the predicted value of the propensity score $P(A_j=a | C,X_{3-j})$ for a twin pair with covariates $C_i,X_{i,3-j}$.  Let $\hat{\mu}_{j,a,b}(C,X_{ij})$ be the predicted outcome regression for Twin $j$ in a twin pair with covariates $C_i,X_{ij}$ if the twins' exposures were set to $A_j=a,A_{3-j}=b$.  Then the efficient estimator for $\beta^{a,b}$ in Model 2, based on data $O_1,\ldots,O_n$, is given by:
\begin{align*} 
\widehat{\beta}^{a,b}_2 =  \ \frac{1}{n} \sum_{i=1}^n  \ \frac{1}{2} \Bigg\{ & \frac{ \mathds{1}(A_{i1}=a,A_{i2}=b)}{ \hat{\pi}_{1,a}(C_i,X_{i1}) \ \hat{\theta}_{2,b}(C_i,X_{i1}  ) }\Big(Y_{i1}- \hat{\mu}_{1,a,b}(C_i,X_{i1}  ) \Big)+ \hat{\mu}_{1,a,b}(C_i,X_{i1} ) \\
& + \ \frac{ \mathds{1}(A_{i1}=b,A_{i2}=a)}{ \hat{\pi}_{2,a}(C_i,X_{i2} ) \ \hat{\theta}_{1 ,b}(C_i,X_{i2})  } \Big(Y_{i2}- \hat{\mu}_{2,a,b}(C_i,X_{i2} ) \Big)+ \hat{\mu}_{2,a,b}(C_i,X_{i2} ) \Big)  \Bigg\}.
\end{align*}
The estimator $\widehat{\beta}_2^{a,b}$ is also doubly robust:  it is consistent if (i) both propensity score models are correct, even if the outcome regression model is incorrect, or (ii) the outcome regression model is correct, even if one or both propensity score models are incorrect.   We show the double robustness of both estimators in Appendix C, and we also illustrate this property in the simulation study in Section 6.  

Efficient influence functions of average main effects and average spillover effects are obtained as differences of efficient influence functions of the $\beta^{a,b}$ parameters.  For example, the efficient influence function of the spillover effect $\beta_{sp}=\beta^{0,0}-\beta^{0,1}$ in Model $k$ is  $\varphi_k(O ; \beta^{0,0})- \varphi_k(O; \beta^{0,1} )$.  Similarly the efficient estimator of $\beta_{sp}$ in Model $k$ is $\widehat{\beta}^{0,0}_k- \widehat{\beta}^{0,1}_k$.

%%%%%%%%%%%%%%%%%%%

\subsection{Confidence intervals}
 
Here we consider confidence intervals for a parameter $\beta^{a,b}= E \big[ Y^{a,b} \big]$ or a linear combination of such parameters, such as an average spillover effect or average main effect, based on the efficient estimators presented above.  

Denote the parameter of interest by $\beta$.  Let $\psi_k(O ; \beta, \nu)$ denote the efficient influence function of $\beta$ in Model $k$,  where $\nu$ represents the parameters of the propensity score models and outcome regression model, which must be estimated in order to obtain the efficient estimator for $\beta$.  Under the assumption that the parameters $\nu$ are estimated consistently at fast enough rates, the resulting estimator for $\beta$ will be asymptotically normal at root-$n$ rates.  Therefore, we may use a Wald-type confidence interval for $\beta$.  Estimating $\nu$ at rates faster than $n^{-1/4}$---that is, using an estimator $\widehat{\nu}$ such that $n^{1/4+\epsilon} \big( \widehat{\nu} - \nu \big)$ is bounded in probability for some $\epsilon >0$---is sufficient.  However, this can be relaxed somewhat since our estimator for $\beta$ is doubly robust, and the bias of a doubly robust estimator is the product of the bias in the propensity score estimation times the bias in the outcome regression estimation.  Therefore it suffices that we estimate the propensity scores and outcome regression at rates whose product is less than $n^{-1/2}$.  Suppose this is the case, and let $\widehat{\beta}_{k,eff}$ be the solution of the estimating equation $\sum_{i=1}^n \psi _k\big(O_i ; \beta, \widehat{\nu} \big) = 0$.  Because of the assumption on the rate of convergence for $\widehat{\nu}$, the influence function of $\widehat{\beta}_{k,eff}$ is the efficient influence function $\psi_k \big( O ; \beta, \nu \big)$ (with the true value of the nuisance parameters $\nu$).  Therefore $\widehat{\beta}_{k,eff}$ is the efficient estimator of $\beta$, and the asymptotic variance of $\widehat{\beta}_{k,eff}$ is the variance of $\psi_k \big( O ; \beta, \nu \big)$. 

Thus in large samples, the variance of $\widehat{\beta}_{k,eff}$ is approximately $\frac{1}{n} Var \big( \psi_k( O ; \beta, \nu ) \big) = \frac{1}{n} E \big[ \psi_k( O ; \beta, \nu)^2 \big]$.  A consistent estimator of $Var \big( \psi_k( O ; \beta, \nu ) \big)$ is $\frac{1}{n} \sum_{i=1}^n \psi _k\big( O_i ;  \widehat{\beta}_{eff}, \hat{\nu} \big)^2$.   Therefore, a large-sample $95\%$ Wald-type confidence interval for $\beta$ is given by:
\[  \widehat{\beta}_{k,eff} \ \pm \ 1.96 \sqrt{ \frac{ \ \sum_{i=1}^n \psi_k \big( O_i; \widehat{\beta}_{k,eff}, \hat{\nu} \big)^2 \ \ }{ n^2 } }  \ . \] 
Alternatively, we may use a bootstrap-based confidence interval, which may have better coverage than the Wald-type confidence interval at moderate sample sizes.

%%%%%%%%%%%%%%%%%%%%%%%%%%%%%%%%%%

\section{Application:  spillover effects of alcohol in early adolescence}

Early use of of alcohol is often associated with substance use and dependence in adulthood \cite{Grant, GrantDawson}.  Recently, Irons et al. \cite{Irons} used a twin design to determine whether this association is due in part to a causal relationship.  The authors found that there is evidence of a causal effect of twins' early alcohol use on their own adult substance use outcomes and antisocial behavior outcomes.  Here we consider the same data from the Minnesota Twin Family Study; but rather than focusing on the effect of twins' exposures on their own outcomes, we investigate whether twins' early alcohol use has a causal impact on their co-twins' outcomes.  That is, our primary question of interest is whether there is evidence of a nonzero spillover effect.  

In the Minnesota Twin Family Study, pairs of same-sex twins were contacted for an intake assessment at which baseline covariates were collected at target age 11;  to assess exposure status at target age 14;  and to measure outcomes at target age 24.  The exposure that we consider here is an indicator of whether the subject had ever consumed alcohol without their parents' permission by the time of the exposure interview.   The outcome on which we focus is Drinking Index, which is a composite measure of adult drinking frequency and amount.  Shared covariates which we include are severity of parent alcohol abuse, severity of parent drug abuse,  parents' occupation level, and sex and zygosity of the twins.  Individual covariates which we include are a measure of academic motivation, number of externalizing disorder symptoms (such as symptoms of attention deficit/hyperactivity disorder or conduct disorder), amount of conflict with parents, and actual age at the time of the exposure assessment.  Here we use only the complete case pairs, which account for 511 of the 761 pairs of twins in the dataset;  future work is needed to develop methods to address missing data in settings where there is interference between pairs.

Our primary target of inference is the spillover effect $\beta_{sp}=E \big[ Y^{0,0}-Y^{0,1} \big]$, which is the mean difference in twins' outcomes we would see if we could intervene to change all co-twins from exposed to unexposed, while keeping all twins unexposed.   We also estimate the main effect $\beta_{main} = E \big[ Y^{0,0} - Y^{1,0} \big]$, the mean difference in twins' outcomes we would see if we could intervene to change all twins from exposed to unexposed, while keeping all co-twins unexposed.  We estimate $\beta_{sp}$ and $\beta_{main}$ using the efficient estimators that we derived in Section 4.   For the Model 1 estimator, we use generalized additive models to model the outcome regression $E[ Y_j | A_j, A_{3-j}, C,X_j,X_{3-j}]$.  In order to model the joint propensity score $P(A_1,A_2 | C,X_1,X_2)$, we first use generalized additive models to fit the  marginal distributions $P(A_1 | C,X_1)$ and $P(A_2 | C,X_2)$, then model the association between the exposures via a Dale model \cite{Dale}.  We allow the strength of the association to vary by sex and zygosity.   For the Model 2 estimator, we use generalized additive models to model the outcome regression $E[ Y_j | A_j, A_{3-j}, C,X_j]$, the propensity score $P(A_j | C,X_j)$, and the propensity score $P(A_j | C,X_{3-j})$.  Details are given in Appendix D.

Results are shown in Table \ref{data_results}.  Using either of the two estimators, there is evidence of a nonzero spillover effect of twins' use of alcohol in early adolescence on their co-twins' adult drinking behavior.  That is, there is evidence of interference between twins in this study.  A spillover effect of  $-1.5$ (the point estimate using the Model 1 estimator) would indicate that twins' Drinking Index outcomes would be an average of 1.5 points lower if all twins and all co-twins were unexposed to alcohol in early adolescence, compared to all twins being unexposed with their co-twins exposed.  The Drinking Index outcome ranges from 0 to 20, with mean 10.7 and standard deviation 4.0.  Larger values correspond to more frequent and/or greater amounts of drinking, so a decrease of 1.5 points would represent a moderate benefit.  There is also evidence of a nonzero main effect.  A main effect of $-2.1$ points would indicate that twins' Drinking Index outcomes would be an average of 2.1 points lower if all twins changed from exposed to unexposed, while their co-twins remained unexposed.  

Our estimators for $\beta_{sp}$ and $\beta_{main}$ are based on the assumption of no unmeasured confounding due to shared or individual factors.  This assumption is untestable, meaning that we have no means of determining that it holds in our study;  and if it does not in fact hold, this could invalidate our results.  However, we can check for one specific way in which the assumption may fail, through comparison of monozygotic (MZ) twins and dizygotic (DZ) twins.   If there is no unmeasured confounding which is due to shared genetics or other unmeasured factors that are differential between MZ and DZ twins, then, since the distribution of measured covariates is well balanced across these two groups, the average spillover effect in the population of MZ twins, $\beta_{sp,MZ}$, should be equal to the average spillover effect in the population of DZ twins, $\beta_{sp,DZ}$.   Therefore, evidence that the average spillover effects are different in these two populations would be evidence that there is unmeasured confounding due to shared genetics.

%%%%%%%%%%%%%%%%%%%%%%%%% Table
\begin{table}

\begin{center}

\begin{tabular}{|c|c|c|c|c|}
\hline 
\  & \ & Spillover effect $ \beta_{sp}$ & Main effect $\beta_{main}$ & CTC effect $\beta_{W}$ \\
\hline 
\hline
All twins & Model 1 &  -1.470 (-2.318, -0.716) &  -2.093 (-3.251, -0.738) & 0.730 (0.064,1.372)   \\
\cline{2-4}
(n=510) &  Model 2 & -1.599 (-2.318, -0.762) & -2.298 (-3.356, -0.823) & \  \\
\hline   \hline
MZ twins & Model 1 & -1.671 (-2.796, -0.640) & -2.131 (-3.422, -1.094) & 0.577 (-0.224, 1.347) \\
\cline{2-4}
(n=324) & Model 2 & -1.931 ( -2.821, -0.705) & -2.424 (-3.499, -1.123) & \  \\
\hline
\hline
DZ twins & Model 1 & -1.039 (-2.523, 0.118) & -1.952 (-3.711, -0.856) & 1.016 (-0.106, 2.155) \\
\cline{2-4}
(n=186) & Model 2 & -0.957 (-2.245, 0.210) & -2.084 (-3.784, -0.890)  & \  \\
\hline
\end{tabular}
\caption{Effect estimates for the MTFS data.  The 3rd and 4th columns show point estimates, and 95\% confidence intervals based on 5,000 bootstrap samples, for the spillover effect $\beta_{sp} = E \big[ Y^{0,0} - Y^{0,1} \big]$ and the main effect $\beta_{main}=E \big[ Y^{0,0} - Y^{1,0} \big]$, using the efficient estimators for Model 1 and for Model 2.  The 5th column shows point estimates, and 95\% confidence intervals based on 5,000 bootstrap samples, for the co-twin control effect $\beta_W=E \big[ Y^{0,1}-Y^{1,0} | A_1 \neq A_2, X_1=X_2 \big]$ estimated by fitting a between-within regression model.  All are shown using data on all twins, using data on MZ twins only, and using data on DZ twins only.}
\label{data_results}
\end{center}
\end{table}
%%%%%%%%%%%%%%%%%%%%%%%%%%%%%%

The point estimates are consistent with a somewhat stronger spillover effect among MZ twins than among DZ twins;  however, a 95\% confidence interval for the parameter $\beta_{sp, MZ} - \beta_{sp,DZ}$ is $(-2.25,1.84)$, so the data do not provide evidence that $\beta_{sp, MZ} - \beta_{sp,DZ}$ is different from zero.  Thus, comparison of the MZ and DZ subgroups does not suggest that there is unmeasured confounding due to shared genetics or factors that are differential between MZ and DZ twins.  However, this subgroup comparison does not provide a means of assessing whether other types of unmeasured confounding may be present.  Unmeasured shared factors such as peer group or attributes of the twins' school or community, and unmeasured individual factors, could still invalidate our results if they are not captured by measured covariates.

We also fit a between-within regression model and report the within-pair coefficient $\beta_W$.  The causal interpretation of $\beta_W$ is valid even if there are shared unmeasured confounders, provided that there is no unmeasured confounding due to individual factors.  In Section 3.1 we saw that, in settings where there is interference between twins, the causal interpretation of $\beta_W$ is equal to the spillover effect minus the main effect on a subset of the group of exposure-discordant twins.  If the estimates of $\beta_W$ were not consistent with the estimates of $\beta_{sp}- \beta_{main}$, this could be an indication that the estimates of $\beta_{sp}$ and $\beta_{main}$ were invalid due to shared confounders (though it could also be an indication that the subgroup effect $\beta_W$ does not generalize to the whole population of twins).  In our case the estimates are consistent with each other.

Irons et al. \cite{Irons} estimated the effect of adolescent alcohol use on adult drinking behavior using two approaches, propensity score weighting and the co-twin control method, and found that the results using the co-twin control method were attenuated compared to the first method.  They pointed out that one possible reason for this attenuation is that unmeasured shared confounders could be creating bias in the propensity score estimates.  Another possible explanation which appears now is interference:  if there is interference between twins in this study, then the parameter estimated using the co-twin control method is not simply the effect of twins' exposure on their outcome, but the difference of this quantity and the spillover effect from their co-twins' exposure.

%%%%%%%%%%%%%%%%%%%%%%%%%%%%%%%%%%%%%%%%%%%

\section{Simulation study}

Here we evaluate the finite-sample performance of the estimators derived in Section 4, using simulated data with sample size $n=500$ designed to mimic the data from the Minnesota Twin Family Study (MTFS).  We consider two data-generating mechanisms.  Under the first data-generating mechanism, the larger Model 1 holds but the smaller Model 2 does not, while under the second data-generating mechanism both models hold. We generate the covariates in the simulated data by resampling covariates from the MTFS dataset.  We then generate the exposures $A_1,A_2$ from a joint distribution of exposures given covariates based on modeling of the MTFS data, and generate the outcomes $Y_1,Y_2$ from a joint distribution of  outcomes given exposures and covariates based on modeling of the MTFS data.   Details of the data-generating mechanisms are given in Appendix D.   

For each data-generating mechanism, we estimate the spillover effect $\beta_{sp}=E \big[Y^{0,0} - Y^{0,1} \big]$ using each of 5 estimators.  Under the first data-generating mechanism, we estimate $\beta_{sp}$ using the efficient estimator for Model 1 with both the joint propensity score model and the outcome regression model correctly specified; the Model 1 estimator with one but not both of these correctly specified;  the Model 1 estimator with both misspecified;  and the Model 2 efficient estimator with the outcome regression and $P(A_j | C,X_j)$ correctly specified.  Table \ref{Dgm1_results} shows the empirical bias of each estimator in $R=5,000$ simulations; the empirical variance of each estimator; the mean of the influence function-based variance estimates;  the mean of the bootstrap variance estimates based on $M=1,000$ bootstrap replicates within each simulation;  the coverage of the 95\% Wald confidence intervals for $\beta_{sp}$ using the influence function-based variance estimates;  and coverage of the percentile bootstrap 95\% confidence intervals.  As expected, bias of the correctly specified Model 1 estimator $\hat{\beta}_1$ is low.  Also as expected, since $\hat{\beta}_1$ is a doubly robust estimator, bias remains low even when either the propensity score model or the outcome regression model is misspecified.  Coverage probabilities of the percentile bootstrap confidence intervals are very close to the nominal 95\% level, while the influence function-based variance estimates are slightly anti-conservative at this sample size, resulting in coverage probabilities that are somewhat lower.

%%%%%%%%%%%%%%%%%%%%%%%%%%%%%% Table
\begin{table}

\begin{center}

\begin{tabular}{|c|c|c|c|c|c|c|}
\hline
\ & Bias & Var & IF-Var Est & Wald Cov'g & Btstp-Var Est & Btstp Cov'g  \\
\hline
$\hat{\beta}^1$ & -0.00658 & 0.11390 & 0.09886 & 92.96 & 0.12582 & 95.12 \\
\hline
$\hat{\beta}^2$ & 0.00320 & 0.11587 &  0.07195 & 87.50 & 0.13357 & 95.68 \\
\hline
$\hat{\beta}_{wr.prop.}^1$ & 0.00042 &  0.11076 & 0.09971 & 93.44 &  0.12522 & 95.10 \\
\hline
$\hat{\beta}_{wr.outc.}^1$ & -0.02243 & 0.11766 & 0.11348 & 94.12 & 0.11136 & 94.70 \\
\hline
$\hat{\beta}_{wr.both}^1$ & -0.04121 & 0.11736 & 0.11618 & 94.56 & 0.11400 & 94.60 \\
\hline

\end{tabular}

\end{center}
\caption{Simulation results for $R=5,000$ simulations at sample size $n=500$ under a data-generating mechanism where Model 1 is correctly specified but Model 2 is not.  Shown are the empirical bias and empirical variance of each estimator; the mean of the $5,000$ influence function-based variance estimates; the mean of the $5,000$ bootstrap-based variance estimates, based on $M=1,000$ bootstraps for each simulation;  and coverage probabilities for the Wald confidence interval which uses the IF-based variance estimate, and for the percentile bootstrap confidence interval.}

\label{Dgm1_results}

\end{table}
%%%%%%%%%%%%%%%%%%%%%%%%%%%%%%%%%%%%%%

%%%%%%%%%%%%%%%%%%%%%%%%%%% Table
\begin{table}

\begin{center}

\begin{tabular}{|c|c|c|c|c|c|c|}
\hline
\ & Bias & Var & IF-Var Est & Wald Cov'g & Btstp-Var Est & Btstp Cov'g  \\
\hline
$\hat{\beta}^1$ & 0.00719 & 0.08526 & 0.07612 & 93.24 & 0.08967 & 95.30 \\
\hline
$\hat{\beta}^2$ & 0.00576 & 0.08273 & 0.07427 & 93.46 & 0.08567 & 95.14  \\
\hline
$\hat{\beta}_{wr.prop.}^2$ & -0.01473 & 0.08232 &  0.07171 & 92.88 & 0.08375 & 94.82  \\
\hline
$\hat{\beta}_{wr.outc.}^2$ & -0.00005 & 0.08710 & 0.08372 & 94.86 & 0.08263 & 95.00 \\
\hline
$\hat{\beta}_{wr.both}^2$ & -0.00453 & 0.08716 & 0.08253 & 94.20 & 0.08231 & 94.50 \\
\hline

\end{tabular}
\caption{Simulation results for $R=5,000$ simulations at sample size $n=500$ under a data-generating mechanism where both Model 1 and Model 2 are correctly specified.  Shown are the empirical bias and empirical variance of each estimator; the mean of the $5,000$ influence function-based variance estimates; the mean of the $5,000$ bootstrap-based variance estimates, based on $M=1,000$ bootstraps for each simulation;  and coverage probabilities for the Wald confidence interval which uses the IF-based variance estimate, and for the percentile bootstrap confidence interval.}

\label{Dgm2_results}

\end{center}
\end{table}
%%%%%%%%%%%%%%%%%%%%%%%%%%%%%%%%%%%%

Under the second data-generating mechanism, we estimate $\beta_{sp}$ using  the efficient estimator for Model 1 with both the joint propensity score model and the outcome regression model correctly specified;  the Model 2 efficient estimator with both propensity score models and the outcome regression model all correctly specified;  the Model 2 estimator with either the propensity score models or the outcome regression model misspecified;  and the Model 2 estimator with all of these misspecified.  Results are displayed in Table \ref{Dgm2_results}.  We expect each of the first four of these estimators to have low bias, as they do.   Coverage probabilities are close to the nominal level, especially using the bootstrap confidence intervals.  

Additionally, the Model 2 efficient estimator has smaller variance than the Model 1 estimator under the second data-generating mechanism.  The theory from Section 4 shows that the Model 2 estimator is asymptotically more efficient than the Model 1 estimator when both models are correct;  and here we see improved precision at sample size $n=500$ in this simulation scenario.

%%%%%%%%%%%%%%%%%%%%%%%%%%%%%%%%%%%

\section{Discussion}

In this paper we have considered the setting of independent pairs of twins where one twin's exposure may have a causal impact on their co-twin's outcome.  Whether or not interference is present in a given study will depend on the nature of the exposure and the outcome:  in some cases, researchers may be able to rule out the possibility of between-twin interference at the outset based on their knowledge of the domain; while for many behavioral exposures and outcomes, interference is likely to be a possibility.  For settings where there may or may not be interference based on scientific considerations, researchers may allow for possible interference and estimate the spillover effect of twins' exposures on their co-twins' outcomes using the estimators we have presented here.  Evidence of a nonzero spillover effect would be evidence of interference.  We have highlighted the impact that between-twin interference would have for researchers using the co-twin control method:  when there is no interference, this method estimates the causal effect of twins' exposures on their outcomes.  When there is interference, however, it estimates the difference of two effects:  the spillover effect of the co-twins' exposures on the twins' outcomes, minus the main effect of twins' exposures on their own outcomes.  

Under the assumption of no unmeasured confounding, we derived the semi-parametric efficient estimators of key causal effects for the interference setting, including average main effects and average spillover effects.  We applied our estimators to data from the Minnesota Twin Family Study (MTFS), and found evidence that twins' exposure to alcohol in early adolescence may have a spillover effect on their co-twins' drinking behavior in adulthood.  However, if there are genetic or other shared or individual unmeasured factors impacting both the choice to drink in early adolescence, and drinking behavior in adulthood (after controlling for measured covariates), this could invalidate our results.  Comparing the causal effects among MZ twin pairs versus among DZ twin pairs did not yield evidence of unmeasured confounders due to shared genetics in the MTFS data.  The development of sensitivity analyses, showing how a range of different strengths of unmeasured confounding would impact results in the interference setting, would be valuable future work.

Future work towards addressing missing data in this setting is also needed.  As with the MTFS data, there may be missingness in baseline covariates, exposures, and outcomes, and the interference structure leads to some challenges in addressing missing data.  An imputation approach, for example, should be designed in a way which is compatible with the different models to be fit for construction of the estimators;  how best to do this for the Model 2 estimator under minimal modeling assumptions is an open research question. 

A final direction of future research involves leveraging symmetry.  Throughout, we randomly labeled the twins in each pair as Twin 1 and Twin 2.  Because this labeling is random, it follows that there is symmetry between the population of twins who are labeled as Twin 1, and the population of twins who are labeled as Twin 2.  In our models we did not explicitly make an assumption of symmetry, and our estimators therefore apply not only to twins data, but to any setting of independent pairs where there is possible interference between partners.  However, in the twins setting, leveraging such a symmetry assumption could lead to additional efficiency gains, and in future work we plan to derive efficient estimators for models which do impose an assumption of symmetry between the twins.

%%%%%%%%%%%%%%%%%%%%%%%%%%%%%%%%%

\section*{Funding}

This work was supported by [ONR grant N00014-18-1-2760 to E.O. and B.S.]; and [R37-AA009367 to M.M.].

\section*{Acknowledgments}
{\it Conflict of Interest}: None declared.

\bibliographystyle{plain}
\bibliography{Twin_paper_bib}

\begin{thebibliography}{10}

\bibitem{Aronow}
Peter~M. Aronow and Cyrus Samii.
\newblock Estimating average causal effects under general interference, with
  application to a social network experiment.
\newblock {\em The Annals of Applied Statistics}, 11(4):1912--1947, 12 2017.

\bibitem{Dale}
Jocelyn~R. Dale.
\newblock Global cross-ratio models for bivariate, discrete, ordered responses.
\newblock {\em Biometrics}, 42(4):909--917, 1986.

\bibitem{GrantDawson}
Bridget~F. Grant and Deborah~A. Dawson.
\newblock Age at onset of alcohol use and its association with dsm-iv alcohol
  abuse and dependence: results from the national longitudinal alcohol
  epidemiologic survey.
\newblock {\em Journal of Substance Abuse}, 9:103 -- 110, 1997.

\bibitem{Grant}
Julia~D. Grant, Jeffrey~F. Scherrer, Michael~T. Lynskey, Michael~J. Lyons,
  Seth~A. Eisen, Ming~T. Tsuang, William~R. True, and Kathleen~K. Bucholz.
\newblock Adolescent alcohol use is a risk factor for adult alcohol and drug
  dependence: evidence from a twin design.
\newblock {\em Psychological Medicine}, 36(1):109--118, 2006.

\bibitem{HongRaudenbush}
Guanglei Hong and Stephen~W Raudenbush.
\newblock Evaluating kindergarten retention policy.
\newblock {\em Journal of the American Statistical Association},
  101(475):901--910, 2006.

\bibitem{HudgensHalloran}
Michael~G Hudgens and M.~Elizabeth Halloran.
\newblock Toward causal inference with interference.
\newblock {\em Journal of the American Statistical Association},
  103(482):832--842, 2008.
\newblock PMID: 19081744.

\bibitem{Irons}
Daniel~E. Irons, William~G. Iacono, and Matt McGue.
\newblock Tests of the effects of adolescent early alcohol exposures on adult
  outcomes.
\newblock {\em Addiction}, 110(2):269--278, 2015.

\bibitem{Johnson}
Wendy Johnson, Eric Turkheimer, Irving~I. Gottesman, and Thomas~J. Bouchard~Jr.
\newblock Beyond heritability: Twin studies in behavioral research.
\newblock {\em Current Directions in Psychological Science}, 18(4):217 -- 220,
  2009.

\bibitem{Kendler}
Kenneth~S. Kendler and Charles~O. Gardner.
\newblock {Dependent Stressful Life Events and Prior Depressive Episodes in the
  Prediction of Major Depression: The Problem of Causal Inference in
  Psychiatric Epidemiology}.
\newblock {\em JAMA Psychiatry}, 67(11):1120--1127, 11 2010.

\bibitem{Lahey}
Benjamin~B. Lahey and Brian~M. D'Onofrio.
\newblock All in the family: Comparing siblings to test causal hypotheses
  regarding environmental influences on behavior.
\newblock {\em Current Directions in Psychological Science}, 19(5):319--323,
  2010.

\bibitem{Lauersen}
Brett Laursen, Amy~C. Hartl, Frank Vitaro, Mara Brendgen, Ginette Dionne, and
  Michel Boivin.
\newblock The spread of substance use and delinquency between adolescent twins.
\newblock {\em Developmental Psychology}, 53(2):329 -- 339, 2017.

\bibitem{LiuIPW}
L.~Liu, M.~G. Hudgens, and S.~Becker-Dreps.
\newblock {On inverse probability-weighted estimators in the presence of
  interference}.
\newblock {\em Biometrika}, 103(4):829--842, 12 2016.

\bibitem{LiuHudgens}
Lan Liu and Michael~G. Hudgens.
\newblock Large sample randomization inference of causal effects in the
  presence of interference.
\newblock {\em Journal of the American Statistical Association},
  109(505):288--301, 2014.
\newblock PMID: 24659836.

\bibitem{LiuDR}
Lan Liu, Michael~G. Hudgens, Bradley Saul, John~D. Clemens, Mohammad Ali, and
  Michael~E. Emch.
\newblock Doubly robust estimation in observational studies with partial
  interference.
\newblock {\em Stat}, 8(1):e214.
\newblock e214 sta4.214.

\bibitem{McGue_utility}
Matt McGue, Merete Osler, and Kaare Christensen.
\newblock Causal inference and observational research: The utility of twins.
\newblock {\em Perspectives on Psychological Science}, 5(5):546--556, 2010.

\bibitem{Miles}
Caleb~H. Miles, Maya Petersen, and Mark~J. van~der Laan.
\newblock Causal inference when counterfactuals depend on the proportion of all
  subjects exposed.
\newblock {\em Biometrics}, 75(3):768--777, 2019.

\bibitem{Oettinger}
Gerald~S. Oettinger.
\newblock Sibling similarity in high school graduation outcomes: Causal
  interdependency or unobserved heterogeneity?.
\newblock {\em Southern Economic Journal}, 66(3):631 -- 648, 2000.

\bibitem{Ogburn}
Elizabeth~L. {Ogburn}, Oleg {Sofrygin}, Ivan {Diaz}, and Mark~J. {van der
  Laan}.
\newblock {Causal inference for social network data}.
\newblock {\em arXiv e-prints}, page arXiv:1705.08527, May 2017.

\bibitem{Schaefer}
Jonathan~D. Schaefer, Terrie~E. Moffitt, Louise Arseneault, Andrea Danese,
  Helen~L. Fisher, Renate Houts, Margaret~A. Sheridan, Jasmin Wertz, and
  Avshalom Caspi.
\newblock Adolescent victimization and early-adult psychopathology: Approaching
  causal inference using a longitudinal twin study to rule out noncausal
  explanations.
\newblock {\em Clinical Psychological Science}, 6(3):352--371, 2018.
\newblock PMID: 29805917.

\bibitem{Scharfstein99}
Daniel~O. Scharfstein, Andrea Rotnitzky, and James~M. Robins.
\newblock Adjusting for nonignorable drop-out using semiparametric nonresponse
  models.
\newblock {\em Journal of the American Statistical Association},
  94(448):1096--1120, 1999.

\bibitem{Sjolander}
Arvid Sj{\"o}lander, Thomas Frisell, and Sara {\"O}–berg.
\newblock Causal interpretation of between-within models for twin research.
\newblock {\em Epidemiologic Methods}, 1(1):217 -- 237, 2012.

\bibitem{Slomkowski}
Cheryl Slomkowski, Richard Rende, Scott Novak, Elizabeth Lloyd-Richardson, and
  Raymond Niaura.
\newblock Sibling effects on smoking in adolescence: evidence for social
  influence from a genetically informative design.
\newblock {\em Addiction}, 100(4):430 -- 438, 2005.

\bibitem{TchetgenVanderWeele}
Eric J~Tchetgen Tchetgen and Tyler~J VanderWeele.
\newblock On causal inference in the presence of interference.
\newblock {\em Statistical Methods in Medical Research}, 21(1):55--75, 2012.
\newblock PMID: 21068053.

\bibitem{AutoG}
Eric~J. {Tchetgen Tchetgen}, Isabel {Fulcher}, and Ilya {Shpitser}.
\newblock {Auto-G-Computation of Causal Effects on a Network}.
\newblock {\em arXiv e-prints}, page arXiv:1709.01577, September 2017.

\bibitem{Tsiatis}
Anastasios~A. Tsiatis.
\newblock {\em Semiparametric Theory and Missing Data}.
\newblock Springer, New York, 2006.

\bibitem{VanDerLaan}
Mark~J. van~der Laan.
\newblock Causal inference for a population of causally connected units.
\newblock {\em Journal of Causal Inference}, 2(1):13 -- 74, 2014.

\end{thebibliography}

%%%%%%%%%%%%%%%%%%%%%%%%%%%%%%%%%%%%%%%%%%%%%%%%%%%%%%%%%%%%%

\appendix

\section{ Identification of the co-twin control effect under interference}

Here we show identification of the subgroup causal effect $\beta_{ctc} := E \big[ Y^{0,1}-Y^{1,0} | A_1 \neq A_2, X_1=X_2=x \big]$ discussed in Section 3.1, under the following assumptions:  
\begin{align*}
\mbox{ \textbf{C1}:}  & \ Y_j^{a,b} \perp \!\!\! \perp (A_1,A_2)  \ | \ (U,X_1,X_2) \\
\mbox{ \textbf{C2}:} & \  \mbox{If } A_j=a, A_{3-j}=b, \mbox{ then } Y_j=Y_j^{ a,b } \\
\mbox{ \textbf{C3}:} &  \mbox{ For all } u,x_1,x_2 \mbox{ in the support of } (U,X_1,X_2), \\
& \ \ \ P \big(A_1=a, A_2=b \ | \ U=u,X_1=x_1,X_2=x_2 \big) >0 \mbox{ for all } a,b=0,1 \\
\mbox{ \textbf{C4}:} & \  \big(U,X_1,X_2, Y_1^{a,b} \big) \mbox{ and } \big(U,X_2,X_1, Y_2^{a,b} \big) \mbox{ are identically distributed}
\end{align*}
\textbf{C1}-\textbf{C3} are untestable assumptions of exchangeability within levels of the (possibly unobserved) shared factors $U$ and the observed individual covariates $X_1,X_2$; consistency; and positivity.  \textbf{C4} is an assumption of symmetry between the population of twins labeled as Twin 1 and the population labeled as Twin 2, which should hold by design as the labeling is randomly assigned.

Let us denote the subgroup of twins who are exposure-discordant with their co-twin, but who have the same value of $x$ of all individual covariates as their co-twin, as $D_x$.  Thus $\beta_{ctc}$ is the mean of the contrast $Y^{0,1}-Y^{1,0}$ over all twins in the subgroup $D_x$.  Note first that this is a weighted average of Twin 1's contrast and Twin 2's contrast in the subgroup of $D_x$ where Twin 1 is exposed and the subgroup where Twin 2 is exposed.  Specifically, writing $\pi := P (A_1=0,A_2=1 , X_1=X_2=x    | A_1 \neq A_2 , X_1=X_2=x)$, we have:

\begin{align*}
 \beta_{ctc}= \frac{1}{2} \bigg\{ & E \big[ Y_1^{0,1}-Y_1^{1,0} | A_1 \neq A_2, X_1=X_2=x \big] + E \big[ Y_2^{0,1}-Y_2^{1,0} | A_1 \neq A_2, X_1=X_2=x \big] \bigg\} \\
  = \ \frac{1}{2}  \bigg\{  & \pi \times E \big[ Y_1^{0,1}- Y_1^{1,0}  | A_1 =0,A_2=1 , X_1=X_2=x \big]  + \\
 & \hspace{.3in} (1-\pi) \times E \big[ Y_1^{0,1}-Y_1^{1,0}  | A_1=1,A_2 =0, X_1=X_2=x \big] \bigg\} + \\
  \ \frac{1}{2}  \bigg\{  & \pi \times E \big[ Y_2^{0,1}- Y_2^{1,0}  | A_1 =0,A_2=1 , X_1=X_2=x \big]   + \\
 &  \hspace{.3in} (1- \pi) \times E \big[ Y_2^{0,1}-Y_2^{1,0}  | A_1=1,A_2 =0, X_1=X_2=x \big]  \bigg\} 
 \end{align*}
Now $E \big[ Y_j^{0,1}  | A_j =0,A_{3-j}=1 , X_1=X_2=x \big]$ is equal to $E \big[ Y_j  | A_j =0,A_{3-j}=1 , X_1=X_2=x \big]$ by consistency, while $E \big[ Y_j^{0,1}  | A_j =1,A_{3-j}=0 , X_1=X_2=x \big]$ can be identified by leveraging the symmetry between the group of twins labeled as Twin 1 and the group labeled as Twin 2:

\begin{align*}
 E \big[ & Y_j^{0,1}  | A_j=1,A_{3-j} =0, X_1=X_2=x \big] \\
  = & \ \int_y y \ dF \big( Y_j^{0,1} = y | A_j=1,A_{3-j}=0,X_1=X_2=x \big) \\
 = & \ \int_u \int_{y} y \ dF \big( Y_j^{0,1} =y | U=u, A_j=1,A_{3-j}=0,X_1=X_2=x \big) \times \\
 & \hspace{1in} dF(u | A_j=1,A_{3-j}=0, X_1=X_2=x )   \\
 = &  \ \int_u \int_{y} y \ dF \big( Y_j^{0,1}=y | U=u, X_1=X_2=x \big) \times  & \mbox{ by \textbf{C1} } \\
 & \hspace{1in} dF \big(u | A_j=1,A_{3-j}=0, X_1=X_2=x \big) \\
 = &  \ \int_u \int_y  y \ dF \big( Y_{3-j}^{0,1}=y | U=u,X_1=X_2=x \big) \times & \mbox{ by \textbf{C4} } \\
 & \hspace{1in} dF \big(u | A_j=1,A_{3-j}=0, X_1=X_2=x \big) \\
 = &  \ \int_u \int_{y} y \ dF \big( Y_{3-j}^{0,1}=y | U=u,A_j=1,A_{3-j}=0, X_1=X_2=x\big) \times   & \mbox{ by \textbf{C1} } \\
 & \hspace{1in} dF \big(u | A_j=1,A_{3-j}=0, X_1=X_2=x \big) \\
 = &  \ \int_u \int_{y} y \ dF \big( Y_{3-j}=y | U=u,A_j=1,A_{3-j}=0, X_1=X_2=x \big)  & \mbox{ by \textbf{C2} } \\
 & \hspace{1in} dF \big(u | A_j=1,A_{3-j}=0, X_1=X_2=x \big) \\
 = & \ E \Big[ Y_{3-j} | A_j=1,A_{3-j}=0, X_1=X_2=x \Big]
 \end{align*}

\noindent Similarly, $E \big[ Y_j^{1,0} | A_j=0,A_{3-j}=1, X_1=X_2=x \big]$ is identified as $E \big[ Y_{3-j} | A_j=0,A_{3-j}=1, X_1=X_2=x \big]$.  Therefore,
\begin{align*}
 \beta_{ctc}=  \ \frac{1}{2}  \bigg\{  & \pi \times E \big[ Y_1- Y_2  | A_1 =0,A_2=1 , X_1=X_2=x \big]  + \\
 & (1-\pi) \times E \big[ Y_2-Y_1  | A_1=1,A_2 =0, X_1=X_2=x \big]  + \\
  \ \frac{1}{2}  \bigg\{  & \pi \times E \big[ Y_1- Y_2  | A_1 =0,A_2=1 , X_1=X_2=x \big]   + \\
 &  (1- \pi) \times E \big[ Y_2-Y_1  | A_1=1,A_2 =0, X_1=X_2=x \big]  \bigg\} \\
 =  E \big[ Y_j & | A_j=1, A_{3-j}=0, X_1=X_2=x \big] - E \big[ Y_j | A_j=0, A_{3-j}=1, X_1=X_2=x \big] .
 \end{align*}

%%%%%%%%%%%%%%%

%%%%%%%%%%%%%%%%%%%%%%%%%%%%%%%%%%%%%%%%%%

\section{ Efficient influence functions }

Here we derive the efficient influence function for the parameter $\beta_j^{a,b} : = E \big[ Y_j^{a,b}  \big]$, for each of the two models presented in Section 3.2.  This also immediately gives the efficient influence function for $\beta^{a,b} = \frac{1}{2} E \big[ Y_1^{a,b} +Y_2^{a,b} \big]$, and of sums and differences of such parameters.  We use the theory, terminology, and notation of Tsiatis \cite{Tsiatis} and Scharfstein et al. \cite{Scharfstein99}.

We assume that we have $n$ independent pairs of twins, and that the data from these twin pairs constitute $n$ i.i.d. draws from some distribution.  Let $O=\big( C,X_1,X_2,A_1,A_2,Y_1,Y_2 \big)$ be the \emph{observed data} for each twin pair.  Let $Z= \big(C,X_1,X_2,A_1,A_2,Y_1^{1,1},Y_2^{1,1},Y_1^{1,0},Y_2^{1,0},Y_1^{0,1}, Y_2^{0,1},Y_1^{0,0}, Y_2^{0,0} \big)$ be the \emph{full data} for each twin pair.  Under the consistency assumption, the observed data is a coarsening of the full data, since $\ds{Y_j= \sum_{a,b=0,1} \mathds{1}(A_j=a,A_{3-j}=b} ) Y_j^{a,b}$ for each $j=1,2$.  We partition the full data $Z$ into $L=\big(C,X_1,X_2, Y_1^{1,1},Y_2^{1,1},Y_1^{1,0},Y_2^{1,0},Y_1^{0,1}, Y_2^{0,1},Y_1^{0,0}, Y_2^{0,0} \big)$ and $R=(A_1,A_2)$, where $R$ determines which components of $Z$ are observed.  

Let $\mathcal{H}^O$ and $\mathcal{H}^Z$ denote the Hilbert spaces of mean-zero functions of the observed data, and of the full data, respectively, with the covariance inner product.    Full-data influence functions for $\beta_j^{a,b}$ are elements of the space $\Lambda^{\perp}$, the orthogonal complement of the full-data nuisance tangent space $\Lambda= \Lambda(F_L) \oplus \Lambda(F_{R|L})$.  Observed-data influence functions are elements of the space $\Lambda^{O,\perp}$, the orthogonal complement of the observed-data nuisance tangent space $\Lambda^O=\Lambda_1^O + \Lambda_2^O$, where $\Lambda_j^O =E[ \Lambda_j |O] := \big\{ g(O) : g(O)= E [ h(Z) | O ]$ for some $h(Z) \in \Lambda_j \big\}$.  The efficient influence function $\varphi(O ; \beta_j^{a,b})$ in a model is the unique influence function for $\beta_j^{a,b}$ which is an element of the observed data tangent space $\mathcal{T}^O$ of that model;  moreover, the efficient influence function is the projection of any influence function onto $\mathcal{T}^O$.

%%%%%%%%%%%%%%%%%%

\subsection{The efficient influence function in Model 1}

Model 1 is the nonparametric model which places no restrictions on the observed data.   This implies that there is just a single observed-data influence function $\varphi_1(O ; \beta_j^{a,b})$ for $\beta_j^{a,b}$ in Model 1, which is therefore the efficient influence function.  By theory in \cite{Scharfstein99} and \cite{Tsiatis}, the observed-data influence function may be found by using a full data influence function for $\beta_j^{a,b}$, say $\varphi^Z \in \Lambda(F_L)$, and the mapping $K: \mathcal{H}^O \to \mathcal{H}^L$ defined by $K( g )= E \big[ \cdot | L \big]$.  An element $h(O)$ in the inverse image of $\varphi^Z$ under this mapping is in the space $\Lambda_1^{O,\perp}$.   This can be seen using adjoint operators:  if we consider the linear map $A: \Lambda_1 \to \mathcal{H}^O$ defined by $A(h)=E[ h | O]$, then $\Lambda_1^{O}$ is the range of $A$.  Therefore, by properties of adjoint operators, $\Lambda_1^{O,\perp}$ is the null space of $A^*$, the adjoint of $A$.  In this case $A^*(g)=E[ g |L] - \Pi \Big( E \big[ g | L \big] | \Lambda_1^{\perp} \Big)$, and therefore $\Lambda_1^{O,\perp}=\big\{ g(O) : E[ g|L ] \in \Lambda_1^\perp \big\}$.  The residual $h(O)- \Pi \big( h(O) | \Lambda_2^O \big)$ from projecting $h(O)$ onto the space $\Lambda_2^O$ is in the space $\Pi( \Lambda_1^{O,\perp} | \Lambda_2^{O,\perp})$.  In our case, the spaces $\Lambda_1^O$ and $\Lambda_2^O$ are orthogonal, which implies that $\Pi( \Lambda_1^{O,\perp} | \Lambda_2^{O,\perp})$ is equal to $\Lambda_1^{O,\perp} \cap \Lambda_2^{O,\perp}=\Lambda^{O,\perp}$.  Therefore the residual is in $\Lambda^{O,\perp}$ and is thus the observed data influence function for $\beta_j^{a,b}$ in Model 1.

A full data influence function for $\beta_j^{a,b}$ is $\varphi^Z = Y_j^{a,b} - \beta_j^{a,b}$.   To verify this, set $\epsilon :=   Y_j^{a,b}- \beta_j^{a,b} $, and partition the parameters of $F_L$ into variation-independent components $\beta_j^{a,b}$ and $\eta$, where $\eta$ is the distribution of $\big( \tilde{L}, \epsilon \big)$ and $\tilde{L}$ is formed by deleting $Y_j^{a,b}$ from the vector $L$.  Then there is a one-to-one transformation from $dF(L)$ to $dF \big( \epsilon, \tilde{L} \big)$, which we factor as $dF( \epsilon) \times dF \big( \tilde{L} | \epsilon \big)$.  Because the only constraint on these distributions is that $\epsilon$ is mean zero, we can show that the nuisance tangent space for $L$ is $\Lambda(F_L)= \Big\{ h \big( \epsilon \big) : E \big[ h(\epsilon) \big] = E \big[ \epsilon h (\epsilon) \big] = 0 \Big\} \oplus \Big\{ h( L) : E \big[ h(L) \ | \ \epsilon \big] = 0 \Big\}$.  Therefore $\epsilon \in \Lambda(F_L)^{\perp}$ as claimed, since $\epsilon$ is orthogonal to each of the direct summands.  

An element of $\mathcal{H}^O$ which is in the inverse image of $\varphi^Z= Y_j^{a,b} - \beta_j^{a,b}$ for the mapping $K$, and hence in $\Lambda_1^{O,\perp}$, is:
\[ g^*(O) = \frac{ \mathds{1}(A_j=a,A_{3-j}=b) }{ P( A_j=a,A_{3-j}=b | C,X_1,X_2) } \Big( Y_j - \beta_j^{a,b} \Big) .\]

\noindent The space $\Lambda(F_{R|L})$ is $\Big\{ h_2(A_1,A_2,C,X_1,X_2 ) : E \big[ h_2(A_1,A_2,C,X_1,X_2 ) | C,X_1,X_2 \big] =0 \Big\}$.  Since elements of $\Lambda(F_{R|L})$ are functions of the observed data (due to the no unobserved confounding assumption), the space $\Lambda(F_{R|L})$ is equal to the space $\Lambda_2^O$.  Projection onto $\Lambda_2^O$ is given by: 
\[ \Pi \big( \cdot | \Lambda_2^O \big) = E \big[ \cdot | A_1,A_2,C,X_1,X_2 \big] - E \big[ \cdot | C,X_1,X_2 \big], \] 
as this operation yields elements that are in $\Lambda_2^O$, and whose residuals are orthogonal to $\Lambda_2^O$.  Below we write $\pi(C,X_1,X_2) :=P(A_j=a,A_{3-j}=b | C,X_1,X_2)$.  Then:

\begin{align*}
  \Pi \big(  g^*(O) | \Lambda_2^O \big)   = & \ \frac{ \mathds{1}{(A_j=a, A_{3-j}=b} ) }{\pi(C,X_1,X_2) } \ E \Big[  Y_j^{a,b} - \beta_j^{a,b}  \big| A_1,A_2,C,X_1,X_2 \Big] \\
& \hspace{.25in} - \frac{1}{ \pi(C,X_1,X_2) } \cdot E \Big[ \mathds{1}{(A_j=a, A_{3-j}=b} ) \big( Y_j^{a,b} - \beta_j^{a,b} \big) \big| C,X_1,X_2 \Big] \  \\
%%%%%%
 =  & \ \frac{ \mathds{1}{(A_j=a, A_{3-j}=b} )}{\pi(C,X_1,X_2) } \ E \left[   Y_j^{a,b} - \beta_j^{a,b}  \ \big| \ C,X_1,X_2 \right] \\
& \hspace{.25in} - \frac{1}{\pi(C,X_1,X_2)} E \Big[ \mathds{1}{(A_j=a, A_{3-j}=b} ) \ \big| \ C,X_1,X_2 \Big] E \Big[  Y_j^{a,b} - \beta_j^{a,b} \ \big| \ C,X_1,X_2 \Big] \  \ \\
%%%%%
=  & \ \frac{ \mathds{1}{(A_j=a, A_{3-j}=b} )}{\pi(C,X_1,X_2) }  E \Big[  Y_j^{a,b} - \beta_j^{a,b}  \ \big| \ A_j=a,A_{3-j}=b,C,X_1,X_2 \Big] \\
& \hspace{.25in} - E \Big[  Y_j^{a,b} - \beta_j^{a,b} \ \big| \ A_j=a,A_{3-j}=b,C,X_1,X_2 \Big] .\\
%%%%%
\end{align*}
Therefore, taking the residual, the efficient influence function for $\beta_j^{a,b}$ for Model 1 is
\begin{align*}
 \varphi_1(O ; \beta_j^{a,b}) = & \ \frac{ \mathds{1}{(A_j=a, A_{3-j}=b} ) }{ P(A_j=a,A_{3-j}=b | C,X_1,X_2) } \Big( Y_j -  E \big[  Y_j | A_j=a,A_{3-j}=b,C,X_1,X_2 \big]  \Big) +  \\
& \hspace{.5in} E \big[  Y_j | A_j=a,A_{3-j}=b,C,X_1,X_2 \big]  - \beta_j^{a,b} . 
\end{align*}

%%%%%%%%%%%%%%%%%%%%%%%%%%%%%%%%%%%%%%%%%%%%%%%%%%%%%%

\subsection{The efficient influence function in Model 2.}

Model 2, by contrast, does impose constraints on the observed data.  In a model that imposes restrictions on the observed law, there will be multiple observed-data influence functions.   An approach for deriving the efficient influence function in such a setting is to (i) find one observed data influence function, say $\varphi_{naive}$, and (ii) project $\varphi_{naive}$ onto the observed data tangent space $\mathcal{T}^O$ for the model.  Here we will actually start by considering a slightly smaller model, which we will call Model 3, in which we impose the additional assumption that  $\big( Y_1^{1,1} , \ldots, Y_1^{0,0} \big) \perp \!\!\! \perp \big( Y_2^{1,1}, \ldots, Y_2^{0,0} \big) | \big( C,X_1,X_2 \big)$.  The observed data tangent space for Model 3 is more straightforward to compute than the observed data tangent space for Model 2;  therefore we will first derive the the efficient influence function for $\beta_j^{a,b}$ in Model 3, then show the efficient influence function for $\beta_j^{a,b}$ in Model 3 is the same as the efficient influence function for $\beta_j^{a,b}$ in Model 2.

We start by recalling the assumptions placed on the full data in Model 3.  Here we write $\tilde{Y}_j$ for the vector of potential outcomes $\big( Y_j^{1,1}, Y_j^{1,0},Y_j^{0,1}, Y_j^{0,0} \big)$:

\begin{align*}
\mbox{ \textbf{A1}:  } & \big( \tilde{Y}_1, \tilde{Y}_2 \big) \perp \!\!\! \perp (A_1,A_2 ) \ \big| \ ( C, X_1,X_2 ) & \mbox{\emph{(exchangeabilitiy)}} \\
\mbox{ \textbf{A2}:  } & \mbox{ For all  } c,x_1,x_2 \mbox{ in the support of } C,X_1,X_2 & \mbox{ \emph{(positivity)}} \\
& \ \ \ P \big(A_1=a,A_2=b \ | \ C=c,X_1=x_1,X_2=x_2 \big) > 0 \mbox{ for all } a,b=0,1 \\
\mbox{\textbf{A3}: } & \mbox{ If } A_1=a, A_2=b, \mbox{ then } Y_1=Y_1^{ a,b }, Y_2=Y_2^{b,a} & \mbox{\emph{(consistency)}} \\
\textbf{A4}:  & \ A_{j} \perp \!\!\! \perp X_{3-j}  \ | \ (C,X_{j} \big) \\
\textbf{A5}:  & \ \tilde{Y}_j  \perp \!\!\! \perp X_{3-j}  \ | \ (C,X_{j} \big) \\
\textbf{A6}:  & \ A_1 \perp \!\!\! \perp A_2 \ | \ (C,X_1,X_2 \big) \\
\textbf{A7}:  & \ \tilde{Y}_1 \perp \!\!\! \perp \tilde{Y}_2  \ | \ (C,X_1,X_2 )
\end{align*}

\noindent Finally, we also assume the property of composition, meaning that for any sets of variables $U,V,W,Z$, if $U \perp \!\!\! \perp V | Z$ and $U \perp \!\!\! \perp W | Z$, then it also holds that $U \perp \!\!\! \perp (V,W) | Z$, as with graphical $d$-separation.  The following independencies follow from the assumptions listed above:

\begin{itemize}

\item \textbf{B1}:  $\tilde{Y}_j \perp \!\!\! \perp (A_1,A_2) \ | \ (C,X_j)$.   
\begin{align*}
dF(a_1,a_2 | \ \tilde{y}_j, c,x_j) = & \ \int_{x_{3-j}} dF( a_1,a_2 | \tilde{y}_j,c,x_j,x_{3-j}) dF(x_{3-j} | \tilde{y}_j ,c,x_j)  \\
= & \  \int_{x_{3-j}} dF( a_1,a_2 | c,x_j,x_{3-j}) dF(x_{3-j} | \tilde{y}_j ,c,x_j) & \textbf{A1} \\
= & \  \int_{x_{3-j}} dF( a_1,a_2 | c,x_j,x_{3-j}) dF( x_{3-j} |c,x_j ) & \textbf{A5} \\
= & \ dF(a_1,a_2 | c,x_j) .
\end{align*}

\item \textbf{B2}: $\tilde{Y}_j \perp \!\!\! \perp X_{3-j}  \ | \ (A_1,A_2,C,X_j)$.  By \textbf{B1}, \textbf{A5}, and composition, we have $\tilde{Y}_j \perp \!\!\! \perp (A_1,A_2,X_{3-j} ) | (C,X_j)$.  Now the result follows immediately from the weak union property of conditional independence, which states that if $U \perp \!\!\! \perp (V,W) | Z$, then $U \perp \!\!\! \perp V | (W,Z)$.

\item \textbf{B3}: $\tilde{Y}_j \perp \!\!\! \perp X_{3-j}  \ | \ (Y_j, A_1,A_2,C,X_j)$.  By \textbf{A3}, for each $a,b=0,1$, $dF \big( x_{3-j}, \tilde{y}_j |  A_j=a,A_{3-j}=b,c,x_j,Y_j=y_j \big) =  dF \big( x_{3-j}, \tilde{y}_j | A_j=a,A_{3-j}=b,c,x_j,Y_j^{a,b}=y_j  \big)$.  Therefore it suffices to show that  $\tilde{Y}_j \perp \!\!\! \perp X_{3-j} | \big(Y_j^{a,b},A_j=a,A_{3-j}=b,C,X_j  \big)$, and this follows immediately from \textbf{B2} by weak union.

\item \textbf{B4}:   $\tilde{Y}_j \perp \!\!\! \perp Y_{3-j} \ | \ (A_1,A_2,C,X_1,X_2)$.  By \textbf{A1}, \textbf{A7}, and composition, $\tilde{Y}_j  \perp \!\!\! \perp (\tilde{Y}_{3-j},A_1,A_2) | (C,X_1,X_2)$.  Since the observed outcome $Y_{3-j}$ is a function of $\tilde{Y}_{3-j},A_1,A_2$, this implies $\tilde{Y}_j  \perp \!\!\! \perp (Y_{3-j},A_1,A_2) | (C,X_1,X_2)$.  Now the result follows by weak union.

\item \textbf{B5}:   $\tilde{Y}_j \perp \!\!\! \perp Y_{3-j} \ | \ (Y_j,A_1,A_2,C,X_1,X_2)$.  By \textbf{B4} and weak union, $\tilde{Y}_j \perp \!\!\! \perp Y_{3-j} | (Y_j^{a,b},A_j=a,A_{3-j}=b,C,X_1,X_2)$.  By \textbf{A3}, this implies $\tilde{Y}_j \perp \!\!\! \perp Y_{3-j} | (Y_j,A_j=a,A_{3-j}=b,C,X_1,X_2)$.

\item \textbf{B6}:  $\tilde{Y}_{j} \perp \!\!\! \perp ( \tilde{Y}_{3-j},X_{3-j}) | (A_1,A_2,C,X_{j})$.

By \textbf{B1}, $\tilde{Y}_j \perp\!\!\! \perp (A_1,A_2) | (C,X_j)$, and by \textbf{A5}, $\tilde{Y}_j \perp \!\!\! \perp X_{3-j} | (C,X_j)$.  Therefore by composition, $\tilde{Y}_j \perp\!\!\! \perp (A_1,A_2,X_{3-j}) | (C,X_j)$.  We show that $\tilde{Y}_j \perp \!\!\! \perp \tilde{Y}_{3-j} \ | \ (C,X_j)$: 
\begin{align*}
dF( \tilde{y}_{3-j} | \ \tilde{y}_j, c,x_j) = & \ \int_{x_{3-j}} dF( \tilde{y}_{3-j} | \tilde{y}_j,c,x_j,x_{3-j}) dF(x_{3-j} | \tilde{y}_j ,c,x_j)  \\
= & \  \int_{x_{3-j}} dF( \tilde{y}_{3-j} | c,x_j,x_{3-j}) dF(x_{3-j} | \tilde{y}_j ,c,x_j) & \textbf{A7} \\
= & \  \int_{x_{3-j}} dF( \tilde{y}_{3-j} | c,x_j,x_{3-j}) dF( x_{3-j} |c,x_j ) & \textbf{A5} \\
= & \ dF(\tilde{y}_{3-j} | c,x_j) .
\end{align*}
Therefore by composition $\tilde{Y}_j \perp \!\!\! \perp  ( \tilde{Y}_{3-j},X_{3-j},A_1,A_2) | (C,X_j)$, and this implies \\ $\tilde{Y}_j \perp \!\!\! \perp ( \tilde{Y}_{3-j},X_{3-j}) | (A_1,A_2,C,X_j)$ by weak union.

\item \textbf{B7}: $A_1 \perp \!\!\! \perp A_2 \ | \ (C,X_j)$.

\begin{align*}
dF(a_{3-j} |  a_j, c,x_j) = & \ \int_{x_{3-j}} dF( a_{3-j} | a_j,c,x_j,x_{3-j}) dF(x_{3-j} | a_j ,c,x_j)  \\
= & \  \int_{x_{3-j}} dF( a_{3-j} | a_j ,c,x_j,x_{3-j}) dF(x_{3-j} | c,x_j) & \textbf{A4} \\
= & \  \int_{x_{3-j}} dF( a_{3-j} | c,x_j,x_{3-j}) dF( x_{3-j} |c,x_j ) & \textbf{A6} \\
= & \ dF(a_{3-j} | c,x_j) .
\end{align*}

\end{itemize}

We now derive the observed data tangent space for Model 3, by first computing the full data tangent space, then showing how to move from the full data to the observed data tangent space.  In Model 3 the full data likelihood factors as:
\begin{align*}
 dF(Z) = & \ dF(c,x_1,x_2) dF \big( \tilde{y}_1,\tilde{y}_2 | c,x_1,x_2 \big) dF \big(a_1,a_2 | \tilde{y}_1,\tilde{y}_2,c,x_1,x_2 \big) & \ \\
 = & \ dF(c,x_1,x_2) dF \big( \tilde{y}_1,\tilde{y}_2 | c,x_1,x_2 \big) dF \big(a_1,a_2 | c,x_1,x_2 \big) & \textbf{A1} \\
 = & \ dF(c,x_1,x_2) dF \big( \tilde{y}_1,\tilde{y}_2 | c,x_1,x_2 \big)   dF(a_1 | c,x_1,x_2)dF(a_2 | c,x_1,x_2) &\textbf{A6}  \\
 = & \ dF(c,x_1,x_2) dF \big( \tilde{y}_1,\tilde{y}_2 | c,x_1,x_2 \big) dF(a_1 | c,x_1) dF(a_2 | c,x_2) & \textbf{A4}  \\
  = & \ dF(c,x_1,x_2) dF \big( \tilde{y}_1 | c,x_1,x_2 \big) dF \big( \tilde{y}_2 | c,x_1,x_2 \big) dF(a_1 | c,x_1) dF(a_2 | c,x_2) & \textbf{A7} \\  
    = & \ dF(c,x_1,x_2) dF \big( \tilde{y}_1 | c,x_1 \big) dF \big( \tilde{y}_2 | c,x_2 \big) dF(a_1 | c,x_1) dF(a_2 | c,x_2).  & \textbf{A5} \\  
 \end{align*}
 
The full data tangent space is $\mathcal{T}= \mathcal{T}_1 \oplus \ldots \oplus \mathcal{T}_5$, where:
\begin{align*}
\mathcal{T}_1= & \  \Big\{ h_1(C,X_1,X_2) : E \big[ h_1(C,X_1,X_2) \big]=0 \Big\}  \\
\mathcal{T}_2 = & \ \Big\{ h_2(\tilde{Y}_1,C,X_1) : E \big[ h_2(\tilde{Y}_1,C,X_1) | C,X_1 \big] =0 \Big\}  \\
\mathcal{T}_3= & \ \Big\{ h_3( \tilde{Y}_2,C,X_2) : E \big[ h_3( \tilde{Y}_2,C,X_2) | C,X_2 \big] =0 \Big\} \\
\mathcal{T}_4 = & \ \Big\{ (A_1-\pi_1) h_4(C,X_1) : h_4(C,X_1) \mbox{ any function } \Big\} \\
\mathcal{T}_5 = & \ \Big\{ (A_2-\pi_2) h_5(C,X_2) : h_5(C,X_2) \mbox{ any function } \Big\}.  
\end{align*}

The observed data tangent space is $\mathcal{T}^O=\mathcal{T}_1^O + \ldots + \mathcal{T}_5^O$, where $\mathcal{T}_k^O= E \big[ \mathcal{T}_k | O \big]$.  For \\ $k=1,4,5$, $\mathcal{T}_k=\mathcal{T}_k^O$, while $\mathcal{T}_2^O = \Big\{ E \big[ h_2( \tilde{Y}_{1}, C, X_1) | O \big] : E[h_2( \tilde{Y}_{1}, C, X_1) | C,X_{1}]=0 \Big\}$ and $\mathcal{T}_3^O = \Big\{ E \big[ h_3( \tilde{Y}_{2}, C, X_2) | O \big] : E[h_3( \tilde{Y}_{2}, C, X_2) | C,X_{2}]=0 \Big\}$.  

We claim that the 5 spaces $\mathcal{T}_k^O, k=1,\ldots,5$ are mutually orthogonal.  To show that $\mathcal{T}_1^O \perp \!\!\! \perp \mathcal{T}_2^O$, let $g_1(C,X_1,X_2) \in \mathcal{T}_1^O$ and $g_2(O)=E [ h_2( \tilde{Y}_{1}, C, X_1) |O] \in \mathcal{T}_2^O$, where $E \big[ h_2( \tilde{Y}_{1}, C, X_1)  | C,X_1 \big]=0$.  Then: 

\begin{align*} 
E \big[ g_1(C,X_1,X_2) g_2(O) \big] = & \ E \Big[ g_1(C,X_1,X_2) E \big[ h_2( \tilde{Y}_{1}, C, X_1) | O \big] \Big] \\
= & \ E \Big[ g_1(C,X_1,X_2) h_2( \tilde{Y}_{1}, C, X_1) \Big]  \\
= & \  E \Big[ g_1(C,X_1,X_2)  E \big[ \underbrace{ h_2( \tilde{Y}_{1}, C, X_1) }_{ \mbox{random in } \tilde{Y}_1 \mbox{ only }} | C,X_1,X_2 \big]  \Big] \\
= & \  E \Big[ g_1(C,X_1,X_2) \underbrace{ E \big[ h_2( \tilde{Y}_{1}, C, X_1) | C,X_1 \big]  }_{=0} \Big] = 0.  & \textbf{A5}
\end{align*}
All other parts of the claim follow similarly from the orthogonality of the full data spaces $\mathcal{T}_1,\ldots,\mathcal{T}_5$, except for showing that $\mathcal{T}_2^O$ and $\mathcal{T}_3^O$ are orthogonal.   To show that $\mathcal{T}_2^O \perp \!\!\! \perp \mathcal{T}_3^O$, let $g_2(O)= E \big[ h_2( \tilde{Y}_1, C, X_1) | O \big] \in \mathcal{T}_2^O$ where $E \big[ h_2( \tilde{Y}_1, C, X_1) | C,X_1 \big]=0$, and let $g_3(O)= E \big[ h_3( \tilde{Y}_2, C, X_2) | O \big] \in \mathcal{T}_3^O$, where $E \big[ h_3( \tilde{Y}_2, C, X_2) | C,X_2 \big] =0$.  Then: 

\begin{align*}
 E & \Big[ g_2(O) g_3(O) \Big] \\
 = & \  E \Big[ E \big[ h_2( \tilde{Y}_1, C, X_1) | O \big] E \big[ h_3( \tilde{Y}_2, C, X_2) | O \big] \Big] \\
 = & \ E \Big[ E \big[ \underbrace{ h_2( \tilde{Y}_1, C, X_1) }_{\mbox{random in } \tilde{Y}_1 \mbox{ only}} | Y_1,Y_2,A_1,A_2,C,X_1,X_2 \big] E \big[ h_3( \tilde{Y}_2, C, X_2) | O \big] \Big] \\
= & \ E \Big[ E \big[ h_2( \tilde{Y}_1, C, X_1) | Y_1,A_1,A_2,C,X_1,X_2 \big] E \big[ h_3( \tilde{Y}_2, C, X_2) | O \big] \Big]  & \textbf{B5}  \\
= & \ E \Big[ E \big[ h_2( \tilde{Y}_1, C, X_1) | Y_1,A_1,A_2,C,X_1 \big] E \big[ h_3( \tilde{Y}_2, C, X_2) | O \big] \Big]  & \textbf{B3}  \\
= & \ E \Big[ E \big[ h_2( \tilde{Y}_1, C, X_1) | Y_1,A_1,A_2,C,X_1 \big]  E \Big[ E \big[ h_3( \tilde{Y}_2, C, X_2) | O \big]  | Y_1,A_1,A_2,C,X_1,X_2 \Big] \Big] \\
= & \ E \Big[ E \big[ h_2( \tilde{Y}_1, C, X_1) | Y_1,A_1,A_2,C,X_1 \big]  E \Big[ \underbrace{h_3( \tilde{Y}_2, C, X_2)}_{\mbox{random in } \tilde{Y}_2 \mbox{ only}}  | Y_1,A_1,A_2,C,X_1,X_2 \Big] \Big] \\
= & \ E \Big[ E \big[ h_2( \tilde{Y}_1, C, X_1) | Y_1,A_1,A_2,C,X_1 \big]  E \Big[ h_3( \tilde{Y}_2, C, X_2)  | A_1,A_2,C,X_1,X_2 \Big] \Big] & \textbf{B4} \\
= & \ E \Big[ E \big[ h_2( \tilde{Y}_1, C, X_1) | Y_1,A_1,A_2,C,X_1 \big]  E \Big[ h_3( \tilde{Y}_2, C, X_2)  | A_1,A_2,C,X_2 \Big] \Big] &  \textbf{B2} \\
= & \ E \Big[ E \big[ h_2( \tilde{Y}_1, C, X_1) | Y_1,A_1,A_2,C,X_1 \big]  \underbrace{ E \Big[ h_3( \tilde{Y}_2, C, X_2)  | C,X_2 \Big] }_{=0}\Big] =0. & \textbf{B1} \\
\end{align*}

Since $\mathcal{T}^O$ is a sum of 5 mutually orthogonal spaces, projection onto $\mathcal{T}^O$ is given by the sum of the projections onto each of the 5 spaces.  Projection onto $\mathcal{T}_1^O$, $\mathcal{T}_4^O$, and $\mathcal{T}_5^O$ is straightforward.  For projection onto $\mathcal{T}_2^O$, we will use the following claim (and analogously for $\mathcal{T}_3^O$):

\underline{Claim:}  $\mathcal{T}_2^O$ is equal to the space $S_2$, where
\[ S_2= \Big\{ h_2(Y_1,A_1,A_2,C,X_1) : E \big[ h_2(Y_1,A_1,A_2,C,X_1) | A_1,A_2,C,X_1 \big] =0 \Big\}.   \] 

To show the claim, we first show that $\mathcal{T}_2^O \subseteq S_2$.
Let $g_2(O)=E \big[ h_2( \tilde{Y}_1, C, X_1) | O \big] \in \mathcal{T}_2^O$.  As shown in the preceding proof, $g_2(O)= E \big[ h_2( \tilde{Y}_1, C, X_1) | Y_1,A_1,A_2,C,X_1 \big]$ by \textbf{B5} and \textbf{B3}, which shows that $g_2(O)$ is a function of $Y_1,A_1,A_2,C,X_1$ only.  We must also check that $E \big[ g_2(O) | A_1,A_2,C,X_1 \big]=0$: 

\begin{align*}
 E \big[ g_2(O) | A_1,A_2,C,X_1 \big] = & \ E \Big[ E \big[ h_2( \tilde{Y}_1,C,X_1) |O \big] | A_1,A_2,C,X_1 \big] \\
 = & \ E \big[ h_2( \tilde{Y}_1, C, X_1) | A_1,A_2,C,X_1 \big] \\
 = & \ E \big[ h( \tilde{Y}_1, C, X_1) | C,X_1 \big]=0.  & \textbf{B1}
 \end{align*}
Therefore $g_2(O) \in S_2$.  To show the opposite containment, we will show that $\mathcal{T}_2^{O,\perp} \subseteq S_2^\perp$, which is equivalent to $\mathcal{T}_2^O \supseteq S_2$.  We begin by computing $\mathcal{T}_2^{O,\perp}$, using properties of adjoint operators.

Consider the linear map $A: \mathcal{T}_2 \to \mathcal{H}^O$ given by $A(h)=E[ h|O]$.  The adjoint of $A$ is the map \\ $A^*: \mathcal{H}^O \to \mathcal{T}_2$ such that $E \big[ A(h_2) \cdot g \big] = E \big[ h_2 \cdot A^*(g) \big]$ for all $h_2 \in \mathcal{T}_2$ and all $g \in \mathcal{H}^O$.  The range of $A$ is $\mathcal{T}_2^O$.  Therefore, by properties of adjoint operators, $\mathcal{T}_2^{O,\perp}$ is the null space of $A^*$.  We show that $A^*$ is given by $A^*(g)=E \big[ g | \tilde{Y}_1,C,X_1 \big] - E \big[ g | C,X_1 \big]$.  First, note that for any $g \in \mathcal{H}^O$, $E \big[ g | \tilde{Y}_1,C,X_1 \big] - E \big[ g | C,X_1 \big]$ is indeed an element in $\mathcal{T}_2$, since it is a function of $\tilde{Y}_1,C,X_1$ that is mean zero given $C,X_1$.  Now we check that $E \big[ A(h_2) g \big] = E \big[ h_2 A^*(g) \big]$:
\begin{align*}
E \Big[ A \Big(h_2(\tilde{Y}_1,C,X_1) \Big) g(O) \Big] = & \ E \Big[ E \big[ h_2(\tilde{Y}_1,C,X_1) | O \big] g(O) \Big] \\
= & \ E \Big[ h_2(\tilde{Y}_1,C,X_1) g(O) \Big] \\
 = & \  E \Big[ h_2(\tilde{Y}_1,C,X_1) E \big[ g(O) | \tilde{Y}_1 ,C,X_1 \big] \Big] \\
= & \ E \Big[ h_2(\tilde{Y}_1,C,X_1) \Big\{ E [ g(O) | \tilde{Y}_1 ,C,X_1] - E[ g(O) | C,X_1] \Big\} \Big]  
\end{align*}
where the last equality follows since 
\[ E \Big[ h_2(\tilde{Y}_1,C,X_1) E[ g(O) | C,X_1] \Big] = E \Big[  \underbrace{E[ h_2(\tilde{Y}_1,C,X_1)  | C,X_1 ] }_{=0} E[ g | C,X_1] \Big] =0  . \]  
Therefore,
\[ \mathcal{T}_2^{O,\perp} = \Big\{ g(O) : E \big[ g | \tilde{Y}_1,C,X_1 \big] = E \big[ g | C,X_1 \big] \Big\} . \]
Now let $W$ be the set
\[ W = \Big\{ g(O) : E \big[ g(O) | \tilde{Y}_1, A_1,A_2,C,X_1 \big] = E \big[ g(O) | A_1,A_2,C,X_1 \big] \Big\}.   \] 
We will show the chain of containments $ \mathcal{T}_2^{O,\perp} \subseteq W \subseteq S_2^{\perp}$.

We first show that, for any $g(O) \in \mathcal{H}^O$, for each $a,b$, $E \big[ g(O) | \tilde{Y}_1, A_1=a,A_2=b,C,X_1 \big]$ is a function of $Y_1^{a,b}$, $C$, and $X_1$ only.  This follows from \textbf{B6}:  since $( \tilde{Y}_2,X_2) \perp \!\!\! \perp \tilde{Y}_1 | (A_1,A_2,C,X_1)$, we can remove all but one of the potential outcomes in $\tilde{Y}_1$ in the conditioning below.
\begin{align*}
E \big[ g(O) | \tilde{Y}_1, & A_1=a,A_2=b,C,X_1 \big] \\
= & \ E \big[ \underbrace{ g \big( Y_1^{a,b}, Y_2^{a,b}, a,b,C,X_1,X_2 \big) }_{ \mbox{random in }Y_2^{a,b}, X_2 \mbox{ only} } | \tilde{Y}_1, A_1=a,A_2=b,C,X_1 \big] \\
= & \ E \big[ g \big( Y_1^{a,b}, Y_2^{a,b}, a,b,C,X_1,X_2 \big) | Y_1^{a,b}, A_1=a,A_2=b,C,X_1 \big]  & \textbf{B6} \\
= : & \ h_{ab}(Y_1^{a,b} ,C,X_1)
\end{align*}

Now by \textbf{B1}, $P(A_1=a,A_2=b | C,X_1, \tilde{Y}_1)=P(A_1=a,A_2=b | C,X_1)$.  Write $\pi_{ab}=\pi_{ab}(C,X_1)$ for this function.  Then we have
\begin{align*}
E  \big[ g(O) | \tilde{Y}_1, C,X_1 \big]  = & \ E \Big[ E \big[ g(O) | \tilde{Y}_1, A_1,A_2,C,X_1 \big] |  \tilde{Y}_1,C,X_1 \Big] \\
= & \ \pi_{11}(C,X_1)  h_{11}(Y_1^{1,1},C,X_1) + \pi_{10}(C,X_1) h_{10}(Y_1^{1,0},C,X_1) + \\
& \hspace{.3in} \pi_{01}(C,X_1) h_{01} (Y_1^{0,1},C,X_1)+\pi_{00}(C,X_1) h_{00}(Y_1^{0,0},C,X_1).
\end{align*}

Now take $g(O) \in \mathcal{T}_2^{O,\perp}$, so that $E \big[ g(O) | \tilde{Y}_1, C,X_1 \big] = E \big[ g(O) | C,X_1 \big]$.  We will show that \\ $E \big[ g(O) | \tilde{Y}_1, A_1,A_2,C,X_1 \big] = E \big[ g(O) | A_1,A_2,C,X_1 \big]$.  Isolating $h_{11}(Y_1^{1,1},C,X_1)$ in the equation \\ $E \big[ g(O) | \tilde{Y}_1, C,X_1 \big] = E \big[ g(O) | C,X_1 \big]$, we have

\[ h(Y_1^{1,1}, C,X_1) = \frac{ E \big[ g(O) | C,X_1 \big] - \pi_{10} h_{10}(Y_1^{1,0}, C,X_1) - \pi_{01}h_{01}(Y_1^{0,1}, C,X_1) - \pi_{00} h_{00}(Y_1^{0,0}, C,X_1) }{ \pi_{11} } .\]

\noindent Since the right hand side does not depend on $Y_1^{1,1}$, this shows that $h(Y_1^{1,1}, C,X_1)$ does not depend on $Y_1^{1,1}$.  Therefore $h(y_1^{1,1}, C,X_1) :=E \big[ g(O) | Y_1^{1,1}=y_1^{1,1}, A_1=1,A_2=1,C,X_1 \big]$ is constant in $y_1^{1,1}$, and hence $E \big[ g(O) | \tilde{Y}_1, A_1=1,A_2=1,C,X_1 \big]= E \big[ g(O) |  A_1=1,A_2=1,C,X_1 \big]$. 

Similarly, $h_{10}$, $h_{0,1}$, and $h_{0,0}$ are functions of $C,X_1$ only, and therefore $E \big[ g(O) | \tilde{Y}_1, A_1=a,A_2=b,C,X_1 \big] = E \big[ g(O) |  A_1=a,A_2=b,C,X_1 \big]$ for each $a,b$.   Therefore $E \big[ g(O) | \tilde{Y}_1, A_1,A_2,C,X_1 \big] = E \big[ g(O) | A_1,A_2,C,X_1 \big]$ as claimed, which shows that $\mathcal{T}_2^{O,\perp} \subseteq W$.

Finally we will show that $W \subseteq S_2^\perp$  Let $g_1(O) \in W$, so that $E \big[ g(O) | \tilde{Y}_1, A_1,A_2,C,X_1 \big] = E \big[ g(O) | A_1,A_2,C,X_1 \big]$.  We show that $g_1(O)$ is orthogonal to every $g_2(Y_1,A_1,A_2,C,X_1) \in S_2$:
\begin{align*}
E  \big[ g_1(O) & g_2(Y_1,A_1,A_2,C,X_1) \big] \\
= & \ E \Big[ E \big[ g_1(O)   | \tilde{Y}_1,A_1,A_2,C,X_1 \big] g_2(Y_1,A_1,A_2,C,X_1) \Big] \\
= & \ E \Big[  E \big[ g_1(O)  | A_1,A_2,C,X_1 \big] g_2(Y_1,A_1,A_2,C,X_1)  \Big]   \\
= & \ E \Big[ E \big[ g_1(O)  | A_1,A_2,C,X_1 \big]  \underbrace{ E \big[ g_2(Y_1,A_1,A_2,C,X_1) | A_1,A_2,C,X_1 \big] }_{=0}   \Big] =0. \\
\end{align*}
This completes the proof of the claim that $\mathcal{T}_2^O = S_2$.  The analogous result holds for $\mathcal{T}_3^O$ by symmetry.  Therefore the observed data tangent space for Model 3 is the direct sum of 5 mutually orthogonal spaces $\mathcal{T}^O = \mathcal{T}_1^O \oplus  \ldots \oplus \mathcal{T}_5^O$, where
\begin{align*}
\mathcal{T}_1^O= & \  \Big\{ h_1(C,X_1,X_2) : E \big[ h_1(C,X_1,X_2) \big]=0 \Big\}  \\
\mathcal{T}_2^O = & \ \Big\{ h_2(Y_1,A_1,A_2,C,X_1) : E \big[ h_2(Y_1,A_1,A_2,C,X_1) | A_1,A_2,C,X_1 \big] =0 \Big\}  \\
\mathcal{T}_3^O= & \ \Big\{ h_3(Y_2,A_1,A_2,,C,X_2) : E \big[ h_3(Y_2,A_1,A_2,C,X_2) | A_1,A_2,C,X_2 \big] =0 \Big\} \\
\mathcal{T}_4^O = & \ \Big\{ (A_1-\pi_1) h_4(C,X_1) : h_4(C,X_1) \mbox{ any function } \Big\} \\
\mathcal{T}_5^O = & \ \Big\{ (A_2-\pi_2) h_5(C,X_2) : h_5(C,X_2) \mbox{ any function } \Big\}.  
\end{align*}

The efficient influence function for $\beta_j^{a,b}$ in Model 3 is the projection of any influence function for $\beta_j^{a,b}$ in Model 3, say $\varphi_3^{naive}$, onto $\mathcal{T}^O$, where by orthogonality, $\Pi \big( \cdot | \mathcal{T}^O \big) = \sum_{k=1}^5 \Pi \big( \cdot | \mathcal{T}_k^O \big)$.  

Because Model 3 is a subset of Model 1, the full-data influence function $\varphi^Z = Y_j^{a,b}- \beta_j^{a,b}$ from Model 1 is also a full-data influence function for Model 3.  Following the steps outlined in Section 1.1,  the element
\[ g^*(O) := \frac{ \mathds{1}(A_j=a,A_{3-j}=b) }{ P( A_j=a,A_{3-j}=b | C,X_1,X_2) } \Big( Y_j - \beta_j^{a,b} \Big) \]
is in the space $\Lambda_1^O$ for Model 3.  In Model 3, $\Lambda_2^O= \Lambda_2 = \mathcal{T}_4^O \oplus \mathcal{T}_5^O$, so one influence function for $\beta_j^{a,b}$ in Model 3 is:
\[ \varphi_3^{naive} := g^*(O) - \Pi \big( g^*(O) | \Lambda_2^O)  = g^*(O) -\Pi \big( g^*(O) | \mathcal{T}_4^O  \big) - \Pi \big( g^*(O) | \mathcal{T}_5^O \big) . \]
Therefore, the efficient influence function for $\beta_j^{a,b}$ in Model 3 is:
\begin{align}
 \varphi_3(O ; \beta_j^{a,b} ) = & \ \Pi \big( \varphi_3^{naive} | \mathcal{T}^O \big) \notag \\ 
 = & \ \Pi \Big(  \big\{ g^*(O) -\Pi \big( g^*(O) | \mathcal{T}_4^O  \big) - \Pi \big( g^*(O) | \mathcal{T}_5^O \big) \big\}  \ \big| \mathcal{T}^O \Big) \notag \\
 = & \  \sum_{k=1}^5  \ \Pi \Big(  \big\{ g^*(O) -\Pi \big( g^*(O) | \mathcal{T}_4^O  \big) - \Pi \big( g^*(O) | \mathcal{T}_5^O \big) \big\}  \ \big| \mathcal{T}_k^O \Big) \notag \\
  = & \ \Pi \big( g^*(O) | \mathcal{T}_1^O \big) +  \Pi \big( g^*(O) | \mathcal{T}_2^O \big) +  \Pi \big( g^*(O) | \mathcal{T}_3^O \big), 
 \end{align}
 where the last line follows because $\Pi \Big( \Pi \big( g^*(O) | \mathcal{T}_k^O \big) | \mathcal{T}_k^O \Big) =  \Pi \big( g^*(O) | \mathcal{T}_k^O \big)$ for each $k$, and because  $\Pi \Big( \Pi \big( g^*(O) | \mathcal{T}_k^O \big) | \mathcal{T}_{k'}^O \Big) = 0$ for any $k \neq k'$, by orthogonality.
 
 For notational convenience we take $a=1,b=1$ in the following calculations;  the calculations for other values of $a,b=0,1$ are exactly similar.  We write $\pi_j=\pi_j(C,X_j) := P(A_j=1 | C,X_j)$, $\pi_{3-j}= \pi_{3-j}(C,X_{3-j}) := P(A_{3-j}=1 | C,X_{3-j})$.  By \textbf{A4} and \textbf{A6}, $P( A_j=1,A_{3-j}=1 | C,X_1,X_2) = \pi_1 \pi_2$. Then: 
 \[ g^*(O) = \frac{ A_1A_2 }{ \pi_1 \pi_2 } \Big( Y_j - \beta_j^{1,1} \Big) =  \frac{ A_1A_2 }{ \pi_1 \pi_2 } \Big( Y_j^{1,1} - \beta_j^{1,1} \Big).  \]
   We compute each of the 3 terms in (1):

 \begin{align*}
  \Pi \Big( g^*(O) | \ \mathcal{T}_1^O \Big) & = E \big[ g^*(O) | C,X_1,X_2 \big] \\
  = & \ E \left[ \frac{ A_1A_2 }{ \pi_1 \pi_2 } \Big( Y_j^{1,1} - \beta_j^{1,1} \Big) \  \big| C,X_1,X_2 \right]  & \ \\
  & = \frac{1}{\pi_1 \pi_2} E \Big[ A_1A_2  \ | \ C,X_1,X_2 \Big]  E \Big[ Y_j^{1,1} - \beta_j^{1,1}  \ | \ C,X_1,X_2 \Big]  &  \textbf{A1} \\
  & = E \Big[ Y_j^{1,1} - \beta_j^{1,1}  \ | \ C,X_1,X_2 \Big]  & \textbf{A4}, \textbf{A6} \\
    & = E \Big[ Y_j^{1,1} - \beta_j^{1,1}  \ | \ C,X_j \Big] & \textbf{A5}  \\
     & = E \Big[ Y_j^{1,1} | A_1=1, A_2=1,C,X_j \Big] - \beta_j^{1,1}   &  \textbf{B1} \\
  & = E \Big[ Y_j | A_1=1, A_2=1,C,X_j \Big] - \beta_j^{1,1}.   &  \textbf{A3}
  \end{align*}
  
We relabel $\mathcal{T}_2^O$ and $\mathcal{T}_3^O$ as $\mathcal{T}_{k^*}^O$ and $\mathcal{T}_{k^\dagger}^O$, according to whether $j=1$ or $j=2$ in the parameter $\beta_j^{a,b}$.  Set $k^* := j+1$, so that $\mathcal{T}_{k*}^O = \Big\{ h_{k^*}(Y_j,A_1,A_2,C,X_j) : E \big[ h_{k^*}(Y_j,A_1,A_2,C,X_j) | A_1,A_2,C,X_j \big] =0 \Big\}$, and $k^\dagger := 4-j$, so $\mathcal{T}_{k^\dagger} ^O = \Big\{ h_{k^\dagger}(Y_{3-j},A_1,A_2,C,X_{3-j}) : E \big[ h_{k^\dagger}(Y_{3-j},A_1,A_2,C,X_{3-j}) | A_1,A_2,C,X_{3-j} \big] =0 \Big\}$.  Then:
 
 \begin{align*}
  \Pi \Big(  g^*(O) & \big| \ \mathcal{T}_{k^*}^O \Big) \\
   =  & \ E \big[ g^*(O) | Y_j,A_1,A_2,C,X_j \big] - E \big[ g^*(O) | A_1,A_2,C,X_j \big] \\
    = & \  \frac{ A_1 A_2}{ \pi_j } \Big(Y_j-\beta_j^{1,1} \Big) E \left[ \frac{1}{\pi_{3-j} } \ \Big| \ Y_j,A_1A_2=1,C,X_j \right]   - \\
    & \hspace{.5in}  \frac{ A_1 A_2}{ \pi_j }  E \left[ \frac{Y_j^{1,1}-\beta_j^{1,1} }{\pi_{3-j} } \ \Big| \ A_1A_2=1,C,X_j \right]  \\
      = & \  \frac{ A_1 A_2}{ \pi_j } \Big(Y_j-\beta_j^{1,1} \Big) E \bigg[ \underbrace{ \frac{1}{\pi_{3-j} (C,X_{3-j})} }_{\mbox{ random in } X_{3-j} \mbox{ only}}  \ \Big| \ Y_j^{1,1},A_1A_2=1,C,X_j \bigg]     & \textbf{A3} \\
    & \hspace{.3in}  - \frac{A_1A_2}{ \pi_j } E \Big[ Y_j^{1,1} - \beta_j^{1,1} \ |  A_1A_2=1,C,X_j \Big] E \left[  \frac{1}{\pi_{3-j} } \Big| A_1A_2=1,C,X_j \right]  & \textbf{B2} \\
     = & \  \frac{ A_1 A_2}{ \pi_j } \Big(Y_j-\beta_j^{1,1} \Big) E \left[ \frac{1}{\pi_{3-j} (C,X_{3-j})}  \ \Big| \ A_1A_2=1,C,X_j \right]     & \textbf{B2} \\
    & \hspace{.3in}  -  \frac{A_1A_2}{ \pi_j } \Big(E \big[ Y_j  |  A_1A_2=1,C,X_j \big]  - \beta_j^{1,1} \Big) E \left[  \frac{1}{\pi_{3-j} } \Big| A_1A_2=1,C,X_j \right]. & \textbf{A3}
  \end{align*}
 \noindent Using the fact that for binary $Z$, $E \big[ ZW | V \big] = E[ W | Z=1,V] P(Z=1 | V)$, we have:
 
\begin{align*}
 E \bigg[ \frac{1}{\pi_{3-j} } & \ \Big| \  A_1A_2=1,C,X_j \bigg] \\
 =  & \ \frac{  1}{ P \big( A_1A_2=1 | C,X_j \big) } E \left[ \frac{A_1A_2}{\pi_{3-j} } \ \big| \ C,X_j \right] \\
 = & \  \frac{ 1 }{ P \big( A_1A_2=1 | C,X_j \big) }  E \left[ \frac{ 1}{\pi_{3-j} }  E \left[  A_1 A_2 \big|  \ C,X_j,X_{3-j} \right]  \ \big| \ C,X_j \right]\\
 = & \  \frac{ 1}{ P(A_1=1,A_2=1 | C,X_j) } E  \left[ \frac{1}{\pi_{3-j} }  ( \pi_j \pi_{3-j} )  \big| \ C,X_j \right]  & \textbf{A4}, \textbf{A6} \\
 = &\  \frac{ 1}{P(A_1=1,A_2=1 | C,X_j) } \pi_j (C,X_j) \\
 = & \  \frac{ 1}{ \pi_j P(A_{3-j}=1 | C,X_j) } \pi_j(C,X_j)  & \textbf{B7} \\
 = & \ \frac{ 1}{ P(A_{3-j}=1 | C,X_j) } .
 \end{align*}
 
\noindent Therefore, 
\[  \Pi \Big( g^*(O) \big| \ \mathcal{T}_{k^*}^O \Big)  = \frac{A_1A_2}{ \pi_j(C,X_j) P(A_{3-j}=1 | C,X_j) } \Big(Y_j - E \big[ Y_j  | A_1A_2=1, C,X_j \big]  \Big)  . \]

Finally,
\begin{align*}
  \Pi \Big(  g^*(O) & \big| \ \mathcal{T}_{k^\dagger}^O \Big) \\
   =  & \ E \big[ g^*(O) | Y_{3-j},A_1,A_2,C,X_{3-j} \big] - E \big[ g^*(O) | A_1,A_2,C,X_{3-j} \big] \\
    = & \  \frac{ A_1 A_2}{ \pi_{3-j} }  E \bigg[ \underbrace{ \ \frac{ Y_j^{1,1}-\beta_j^{1,1} }{\pi_j(C,X_j) }  \ }_{ \mbox{random in } Y_j^{1,1}, X_j \mbox{ only} } \ \Big| \ Y_{3-j}^{1,1},A_1A_2=1,C,X_{3-j} \bigg]   \\
    & \hspace{.5in} - \frac{ A_1 A_2}{ \pi_j }  E \left[ \frac{Y_j^{1,1}-\beta_j^{1,1} }{\pi_{3-j} } \ \Big| \ A_1A_2=1,C,X_{3-j} \right]  \\
      = & \  \frac{ A_1 A_2}{ \pi_{3-j} }  E \bigg[  \frac{ Y_j^{1,1}-\beta_j^{1,1} }{\pi_j(C,X_j) }   \ \Big| \ A_1A_2=1,C,X_{3-j} \bigg]   & \textbf{B6} \\
    & \hspace{.5in} - \frac{ A_1 A_2}{ \pi_j }  E \left[ \frac{Y_j^{1,1}-\beta_j^{1,1} }{\pi_{3-j} } \ \Big| \ A_1A_2=1,C,X_{3-j} \right]  \\
    =& \ 0.
  \end{align*}

\noindent  Therefore, the efficient influence function for $\beta_j^{1,1}$ in Model 3 is:
\begin{align*}
 \varphi_3 (  O ; \beta_j^{1,1} ) = & \ \Pi \big( g^*(O) | \mathcal{T}_1^O \big) +  \Pi \big( g^*(O) | \mathcal{T}_{k^*}^O \big) +  \Pi \big( g^*(O) | \mathcal{T}_{k^\dagger}^O \big) \\
 = & \  \frac{A_1A_2}{ \pi_j(C,X_j) P(A_{3-j}=1 | C,X_j) } \Big(Y_j - E \big[ Y_j  | A_1A_2=1, C,X_j \big]  \Big) \\
& \hspace{.3in} +  E \Big[ Y_j | A_1=1, A_2=1,C,X_j \Big] - \beta_j^{1,1} .
  \end{align*}
  By exactly the same arguments, the efficient influence function for $\beta_j^{a,b}$ in Model 3 is:
  \begin{align*}
 \varphi_3 (  O ; \beta_j^{a,b} ) =  & \  \frac{ \mathds{1}(A_j=a,A_{3-j}=b)}{ P(A_j=a | C,X_j) P(A_{3-j}=b | C,X_j) } \Big(Y_j - E \big[ Y_j  | A_j=a,A_{3-j}=b, C,X_j \big]  \Big) \\
& \hspace{.3in} +  E \Big[ Y_j | A_j=a, A_{3-j}=b,C,X_j \Big] - \beta_j^{a,b} .
  \end{align*}

Next we will show that the function $\varphi_3 (  O ; \beta_j^{a,b} )$ given above is also the efficient influence function for $\beta_j^{a,b}$ in Model 2.  It suffices to show that $\varphi_3(O; \beta_j^{a,b} )$ is an influence function for $\beta_j^{a,b}$ in Model 2, for then the projection of $\varphi_3(O ; \beta_j^{a,b})$ onto $\mathcal{T}^O(M2)$ is the efficient influence function for $\beta_j^{a,b}$ in Model 2.  But Model 3 $\subseteq$ Model 2, which implies $\mathcal{T}^O(M3) \subseteq \mathcal{T}^O(M2)$.   Since $\varphi_3 (  O ; \beta_j^{a,b} )$ is the efficient influence function for Model 3, we know that $\varphi_3 (  O ; \beta_j^{a,b} ) \in \mathcal{T}^O(M3)$.  Therefore $\varphi_3(O ; \beta_j^{a,b}) \in \mathcal{T}^O(M2)$, and so is its own projection onto $\mathcal{T}^O(M2)$.  Thus, once we show that $\varphi_3(O; \beta_j^{a,b} )$ is an influence function for $\beta_j^{a,b}$ in Model 2, it will automatically follow that $\varphi_3(O; \beta_j^{a,b} )$ is the efficient influence function for $\beta_j^{a,b}$ in Model 2.

To show that $\varphi_3(O ; \beta_j^{a,b})$ is an influence function for $\beta_j^{a,b}$ in Model 2, let $\mathcal{P}$ be any regular parametric submodel of Model 2, parametrized say by $\gamma \in \Gamma_{\mathcal{P}} \subseteq \mathbb{R}^r$, in such a way that $\gamma=0$ corresponds to the true distribution $P_0 \in \mathcal{P}$.  Write $\mathcal{P} = \Big\{ f_{\mathcal{P}}(O; \gamma) :  \gamma \in \Gamma_{\mathcal{P}} \Big\}$, and let $\ds{ S_\gamma(O; \mathcal{P} )=  \frac{ \partial \log f_{\mathcal{P} } ( O ; \gamma) }{ \partial \gamma}  \Big|_{\gamma=0} }$ be the score vector for $\mathcal{P}$ evaluated at the truth. We will show that:

\begin{equation} 
E \bigg[   \varphi_3(O ; \beta_j^{a,b} ) \ S_{\gamma}(O ; \mathcal{P} )  \Big] \ = \ \frac{ \partial \beta_j^{a,b}( \mathcal{P} ; \gamma) }{\partial \gamma} \Big|_{\gamma=0 } 
\end{equation}

\medskip

\noindent where the right hand side is the derivative of the functional $\beta_j^{a,b}=\beta_j^{a,b}( \mathcal{P} ; \gamma)$, evaluated at the truth.    By \cite{Tsiatis},  $\varphi_3(O; \beta_j^{a,b})$ is an influence function for $\beta_j^{a,b}$ in Model 2 if and only if equation (2) is satisfied for every regular parametric submodel of Model 2.

Factoring the observed data likelihood, $\mathcal{P}$ can be parametrized as
\[ \hspace{-1in} \mathcal{P} = \Big\{ dF( c,x_1,x_2 ,a_1,a_2 \ ; \gamma_1) \cdot dF(y_j | c,x_j,a_1,a_2 \ ; \gamma_2 ) \cdot \]
\[ \hspace{1in}  dF( y_{3-j} | c,x_1,x_2,a_1,a_2, y_j  \ ; \gamma_3 ) :  \gamma= (\gamma_1,\gamma_2,\gamma_3) \in \Gamma_1 \times \Gamma_2 \times \Gamma_3 = \Gamma_{\mathcal{P}} \Big\} \]

\noindent where $\gamma_1,\gamma_2,\gamma_3$ are variation independent.  We can partition the set of equations (B.2) into the 3 sets of equations
\begin{equation}
 E \bigg[   \varphi_3(O ; \beta_j^{a,b} ) \ S_{\gamma_k}(O ; \mathcal{P} )  \Big] \ = \ \frac{ \partial \beta_j^{a,b}( \mathcal{P} ; \gamma ) }{\partial \gamma_k} \Big|_{\gamma_k=0 } 
 \end{equation}
for $k=1,2,3$.  We will show that, for $k=3$, both sides of (3) are zero, since neither the functional $\beta_j^{a,b}( \mathcal{P} ; \gamma)$ nor the the influence function $\varphi_3(O ; \beta_j^{a,b} )$ involve $Y_{3-j}$.   We will then show that (3) holds for $k=1,2$ because Model 2 and Model 3 are identical as regards features related to $\gamma_1$ and $\gamma_2$.

The identifying functional for $\beta_j^{a,b}( \mathcal{P} ; \gamma)$ is:

\begin{equation}  
\beta_j^{a,b} ( \mathcal{P} ; \gamma) =  \int_{c,x_1,x_2} \int_{y_j}   y_j \cdot dF \big( y_j | c,x_j,a_j=a,a_{3-j}=b \ ; \gamma_2 \big)  dF \big( c,x_1,x_2 \ ; \gamma_1 \big) .
\end{equation}

\medskip

\noindent Note that (4) varies in $\gamma_1,\gamma_2$ only, so  $\ds{ \frac{ \partial \beta_j^{a,b}( \mathcal{P} ; \gamma) }{\partial \gamma_3} =0 }$.  Since $S_{\gamma_3}(O ; \mathcal{P} )$ is the score for \\ $dF( y_{3-j} | c,x_1,x_2,a_1,a_2, y_1 )$, $E \Big[ S_{\gamma_3}(O ; \mathcal{P} ) | C,X_1,X_2,A_1,A_2,Y_j \Big] =0$.  Note also that $\varphi_3(O ; \beta_j^{a,b} )$ is a function of $Y_j,A_1,A_2,C,X_j$ only.  Therefore:

\[ E \bigg[   \varphi_3(O ; \beta_j^{a,b} ) \ S_{\gamma_3}(O ; \mathcal{P} )  \Big] = E \bigg[   \varphi_3(O ; \beta_j^{a,b} )  \underbrace{ E \Big[ S_{\gamma_3}(O ; \mathcal{P} ) | C,X_1,X_2,A_1,A_2,Y_j \Big] }_{=0} \bigg] = 0 \]
which shows that (3) is satisfied for $k=3$.

Now consider the submodel $\mathcal{P}' \subseteq \mathcal{P}$ given by
\[  \hspace{-1in} \mathcal{P}' := \Big\{ dF( c,x_1,x_2,a_1,a_2 ; \gamma_1) \cdot dF(y_j | c,x_j,a_1,a_2 ; \gamma_2 )  \cdot  \]
\[ \hspace{1in} dF_0(y_{3-j} | c,x_1,x_2,a_1,a_2,y_j; \gamma_3=0) :  \gamma'=(\gamma_1,\gamma_2,0 ) \in \Gamma_1 \times \Gamma_2 \times \{0 \}  \Big\} \]
Note that for $k=1,2$, $\mathcal{P}$ and $\mathcal{P}'$ have the same scores, i.e. $S_{\gamma_k}(O ; \mathcal{P} )= S_{\gamma_k}(O ; \mathcal{P}' )$, and that $\beta_j^{a,b} (  \mathcal{P} ; \gamma)= \beta_j^{a,b} (  \mathcal{P}' ; \gamma')$.  We claim that $\mathcal{P}'$ is a parametric submodel of Model 3:  Model 2 and Model 3 impose exactly the same restrictions on $dF(c,x_1,x_2,a_1,a_2)$, and neither imposes any restriction on $dF(y_j | c,x_j,a_1,a_2)$.   For each $(\gamma_1,\gamma_2) \in \Gamma_1 \times \Gamma_2$,  $dF( c,x_1,x_2,a_1,a_2 ; \gamma_1) \cdot dF(y_j | c,x_j,a_1,a_2 ; \gamma_2 )  \cdot dF_0(y_{3-j} | c,x_1,x_2,a_1,a_2,y_j; \gamma_3=0)$ is in Model 2;  therefore it is also in Model 3.  Furthermore, taking $\gamma_1=\gamma_2=0$ shows that $\mathcal{P}'$ contains the true distribution.  

Therefore, since $\varphi_3(O ; \beta_j^{a,b} )$ is an influence function for $\beta_j^{a,b}$ in Model 3, and $\mathcal{P}'$ is a regular parametric submodel of Model 3, 
\[  E \bigg[   \varphi_3(O ; \beta_j^{a,b} ) \ S_{\gamma_k}(O ; \mathcal{P}' )  \Big] \ = \ \frac{ \partial \beta_j^{a,b}( \mathcal{P}' ; \gamma ) }{\partial \gamma_k} \Big|_{\gamma_k=0 } .\]
Since $S_{\gamma_k}(O ; \mathcal{P} )= S_{\gamma_k}(O ; \mathcal{P}' )$ for $k=1,2$ and $\beta_j^{a,b} (  \mathcal{P} ; \gamma)= \beta_j^{a,b} (  \mathcal{P}' ; \gamma')$, this shows that equation (3) holds for $k=1,2$ as claimed.   Therefore,  $\varphi_3(O ; \beta_j^{a,b} )$ is an influence function for $\beta_j^{a,b}$ in Model 2, and hence the efficient influence function for $\beta_j^{a,b}$ in Model 2.

%%%%%%%%%%%%%%%%%%%%%%%%%%%%%%%%%%%%%%%%%%

\section{Double Robustness}

Here we show that the Model 1 and Model 2 efficient estimators for $\beta_j^{a,b} = E \big[ Y_j^{a,b} \big]$ are doubly robust.  

We consider the Model 1 estimator first.  Fix $j$, $a$, and $b$, and let $\pi(C,X_j,X_{3-j} ; \alpha)$ be a model for the propensity score $P(A_j=a,A_{3-j}=b | C,X_j,X_{3-j})$, and let $\mu(A_j,A_{3-j},C,X_j,X_{3-j} ; \gamma)$ be a model for $E \big[ Y_j | A_j,A_{3-j},C,X_j,X_{3-j} \big]$.  These models may or may not be correctly specified, but suppose that they converge to some values, say $\hat{\alpha} \overset{p}{\to} \alpha^*$ and $\hat{\gamma} \overset{p}{\to} \gamma^*$.  In the case where the models are correctly specified, say with $\alpha_0$ and $\gamma_0$ corresponding to the truth, we would have $\alpha^*=\alpha_0$ and $\gamma^*= \gamma_0$.  Let $\pi(C_i,X_{ij}, X_{i,3-j} ; \hat{\alpha})$ and $\mu( a,b, C_i,X_{ij}, X_{i,3-j} ; \hat{\gamma})$ denote predicted values under these models, and consider the corresponding Model 1 estimator
 \begin{align*} 
 \widehat{\beta}_1 =  \ \frac{1}{n} \sum_{i=1}^n   \Bigg\{ & \frac{ \mathds{1}(A_{ij}=a,A_{i,3-j}=b)}{ \pi( C_i,X_{ij},X_{i,3-j} ; \hat{\alpha}) }\Big[Y_{ij}- \mu \big(a,b,C_i,X_{ij},X_{i,3-j} ; \hat{\gamma} \big) \Big] \\ 
 & \hspace{1in} +  \mu \big(C_i,X_{ij},X_{i,3-j} ; \hat{\gamma} \big)  \Bigg\} .
\end{align*}
We will show that $\hat{\beta}_1$ is consistent even if either $\pi$ or $\mu$ is misspecified, as long as the other is correctly specified.  We claim that $\hat{\beta}_1 \overset{p}{\longrightarrow} \beta^*$, where
\begin{align*}
 \beta^* =  \ E  & \Bigg[ \ \frac{\mathds{1}(A_j=a,A_{3-j}=b) }{ \pi(C,X_j,X_{3-j} ; \alpha^* )} \Big( Y_j - \mu(a,b,C,X_j,X_{3-j} ; \gamma^*) \Big) + \mu(a,b,C,X_j,X_{3-j}; \gamma^* )  \Bigg].
 \end{align*}
 For we can rewrite $\hat{\beta}_1$ as
 \begin{equation}  \frac{1}{n} \sum_{i=1}^n   \Bigg\{  \frac{ \mathds{1}(A_{ij}=a,A_{i,3-j}=b)}{ \pi( C_i,X_{ij},X_{i,3-j} ; \alpha^* ) } \Big(Y_{ij}- \mu(a,b,C_i,X_{ij},X_{i,3-j} ; \gamma^* ) \Big)+ \mu(a,b,C_i,X_{ij},X_{i,3-j} ; \gamma^* ) \Bigg\} 
 \end{equation}
 \begin{align} 
 +\ \  \frac{1}{n} \sum_{i=1}^n \Bigg\{   & \left[  \frac{ \pi(C_i,X_{ij},X_{i,3-j}; \alpha^*) - \pi(C_i,X_{ij},X_{i,3-j} ; \hat{\alpha}) }{ \pi(C_i,X_{ij},X_{i,3-j} ; \alpha^*)  \pi(C_i,X_{ij},X_{i,3-j} ; \hat{\alpha}) } \right]  \mathds{1}(A_{ij}=a,A_{i,3-j}=b) \times \notag \\
 & \hspace{1in} \Big[ Y_{ij} - \mu(a,b,C_i,X_{ij},X_{i,3-j} ; \hat{\gamma}) \Big]  \\
 + & \  \left[1 - \frac{ \mathds{1}(A_{ij}=a,A_{i,3-j}=b)}{ \pi(C_i,X_{ij},X_{i,3-j} ; \alpha^*) } \right] \Big[ \mu(a,b,C_i,X_{ij},X_{i,3-j} ; \hat{\gamma}) - \mu(a,b,C_i,X_{ij},X_{i,3-j} ; \gamma^*) \Big]  \Bigg\} \notag
 \end{align}
 where (5) is a sample mean of i.i.d. terms, and hence converges to the expected value of each term, and (6) converges in probability to 0.

 Now we show that, if either $\pi$ or $\mu$ is correctly specified, then $\beta^* = E \big[ Y_j^{a,b} \big]$.  If $\mu$ is correctly specified, then $\mu(a,b,C,X_j,X_{3-j} ; \gamma^*)= E \big[ Y_j^{a,b} | C, X_j, X_{3-j} \big]$.  Conditioning on $Y_j^{a,b}, C,X_j,X_{3-j}$, we have:

\begin{align*}
\beta^* = & \ E \Bigg[ \frac{  Y_j^{a,b} - E \big[ Y_j^{a,b} |  C,X_j,X_{3-j}  \big]}{ \pi(C,X_j,X_{3-j} ; \alpha^* )}  E \Big[ \mathds{1}(A_j=a,A_{3-j}=b) \ \big| Y_j^{a,b},C, X_j,X_{3-j} \Big]   \\ 
& \hspace{1in} + E \big[ Y_j^{a,b} | C, X_j, X_{3-j} \big] \Bigg]  \\
= & \ E \Bigg[ \frac{  Y_j^{a,b} - E \big[ Y_j^{a,b} |  C,X_j,X_{3-j}  \big]}{ \pi(C,X_j,X_{3-j} ; \alpha^* )}  E \Big[ \mathds{1}(A_j=a,A_{3-j}=b) \ \big| C, X_j,X_{3-j} \Big]  & \mbox{ by \textbf{A1} } \\ 
& \hspace{1in} + E \big[ Y_j^{a,b} | C, X_j, X_{3-j} \big] \Bigg]   \\
 = & \ E \Bigg[ \frac{ P(A_j=a,A_{3-j}=b) | C, X_j,X_{3-j} ) }{ \pi(C,X_j,X_{3-j}  ; \alpha^*)}  \underbrace{ E \Big[  \Big( Y_j^{a,b} - E \big[ Y_j^{a,b} |  C,X_j,X_{3-j}  \big]  \Big) \big| C,X_j,X_{3-j} \Big] }_{=0}  \\
 & \hspace{1in} + E \big[ Y_j^{a,b} | C, X_1, X_2 \big]  \Bigg] \\
 = & \  0 + E \big[ Y_j^{a,b} \big] .
 \end{align*}
 If $\pi$ is correctly specified, then we have:
\begin{align*}
\beta^* = & \ E \Bigg[ \frac{  Y_j^{a,b} - \mu(a,b,C,X_j,X_{3-j} ; \gamma^* ) }{ \pi(C,X_j,X_{3-j} ; \alpha_0)}  E \Big[ \mathds{1}(A_j=a,A_{3-j}=b) | Y_j^{a,b},C, X_j,X_{3-j} \Big]  \\
& \hspace{1in} + \mu(a,b, C, X_j, X_{3-j} ; \gamma^* ) \Bigg] \\
 = & \ E \Bigg[ \frac{  Y_j^{a,b} - \mu(a,b,C,X_j,X_{3-j} ; \gamma^* ) }{ \pi(C,X_j,X_{3-j} ; \alpha_0)}  E \Big[ \mathds{1}(A_j=a,A_{3-j}=b) | C, X_j,X_{3-j} \Big]  & \mbox{ by \textbf{A1} } \\
& \hspace{1in} + \mu(a,b, C, X_j, X_{3-j} ; \gamma^* ) \Bigg] \\
   = & \ E \Big[  Y_j^{a,b} -\mu(a,b,  C,X_1,X_2 ; \gamma^*  )   + \mu(a,b, C, X_1, X_2 ; \gamma^* ) \Big] \\
   = & \  E \big[ Y_j^{a,b} \big] .
 \end{align*}

%%%%%%%%%%%%%%%%%%%%%%%%%%%%

Next we consider the Model 2 estimator.  Now let $\pi(C,X_j ; \alpha)$ be a model for the propensity score $P(A_j=a | C,X_j)$, let $\psi( C, X_{j} ; \theta)$ be a model for the propensity score $P(A_{3-j}=b | C, X_j)$, and let $\mu(A_j,A_{3-j},C,X_j; \gamma)$ be a model for $E \big[ Y_j | A_j,A_{3-j},C,X_j \big]$.  Suppose that these models converge to some values, say $\hat{\alpha} \overset{p}{\to} \alpha^*$, $\hat{\theta} \overset{p}{\to} \theta^*$, and $\hat{\gamma} \overset{p}{\to} \gamma^*$.  In the case where the models are correctly specified, denote the truth by $\alpha_0$, $\theta_0$, and $\gamma_0$ respectively.  Let $\pi(C_i,X_{ij} ; \hat{\alpha})$, $\psi(C_i,X_{ij} ; \hat{\theta})$, and $\mu( a,b, C_i,X_{ij} ; \hat{\gamma})$ denote predicted values under these models, and consider the corresponding Model 2 estimator:
\[ \widehat{\beta}_2 =  \ \frac{1}{n} \sum_{i=1}^n  \Bigg\{  \frac{ \mathds{1}(A_{ij}=a,A_{i,3-j}=b)}{ \pi( C_i,X_{ij}; \hat{\alpha}) \psi(C_i,X_{ij} ; \hat{\theta} ) } \Big(Y_{ij}- \mu(a,b,C_i,X_{ij} ; \hat{\gamma} ) \Big)+ \mu(a,b,C_i,X_{ij}  ; \hat{\gamma} ) \Bigg\} . \]
We show that $\hat{\beta}_2$ is consistent under misspecification of one or both of the propensity score models, provided the outcome regression model is correctly specified;  and that $\hat{\beta}_2$ is consistent under misspecification of the outcome regression model, provided that both propensity score models are correctly specified.  

We have $\hat{\beta}_2 \overset{p}{\longrightarrow} \beta^*$, where
\[  \beta^* =\ E   \Bigg[ \ \frac{\mathds{1}(A_j=a,A_{3-j}=b) }{ \pi(C,X_j ; \alpha^* ) \ \psi(C,X_j ; \theta^*)} \Big( Y_j - \mu(a,b,C,X_j ; \gamma^* \Big) + \mu(a,b, C,X_j ; \gamma^* )   \Bigg]. \]
Now suppose first that the outcome regression model is correctly specified.  Then $\mu(a,b,C,X_j ; \gamma^*)= E \big[ Y_j^{a,b} | C, X_j \big]$.  Conditioning on $Y_j^{a,b}, C,X_j$, we have:
\begin{align*}
\beta^* = & \ E \Bigg[ \frac{  Y_j^{a,b} - E \big[ Y_j^{a,b} |  C,X_j  \big]}{ \pi(C,X_j; \alpha^* ) \psi(C,X_j ; \theta^*) }  E \Big[ \mathds{1}(A_j=a,A_{3-j}=b) | Y_j^{a,b},C, X_j \Big]  + E \big[ Y_j^{a,b} | C, X_j \big] \Bigg]  \\
= & \ E \Bigg[ \frac{  Y_j^{a,b} - E \big[ Y_j^{a,b} |  C,X_j  \big]}{ \pi(C,X_j; \alpha^* ) \psi(C,X_j ; \theta^*) }  E \Big[ \mathds{1}(A_j=a,A_{3-j}=b) | C, X_j \Big]  + E \big[ Y_j^{a,b} | C, X_j \big] \Bigg] & \mbox{ by \textbf{B1}} \\
 = & \ E \Bigg[ \frac{ P(A_j=a,A_{3-j}=b) | C, X_j ) }{ \pi(C,X_j; \alpha^* ) \ \psi(C,X_j ; \theta^*) }  \underbrace{ E \Big[  \Big( Y_j^{a,b} - E \big[ Y_j^{a,b} |  C,X_j  \big]  \Big) \big| C,X_j \Big] }_{=0} + E \big[ Y_j^{a,b} | C, X_j \big]  \Bigg] \\
 = & \  0 + E \big[ Y_j^{a,b} \big] .
 \end{align*}
Finally, if both propensity score models are correctly specified, then conditioning on $Y_j^{a,b},C,X_j$, we have:

\begin{align*}
\beta^* = & \ E \Bigg[ \frac{  Y_j^{a,b} - \mu(a,b,C,X_j  ; \gamma^*) }{ \pi(C,X_j ; \alpha_0 ) \psi(C,X_j ; \theta_0) }  E \Big[ \mathds{1}(A_j=a,A_{3-j}=b) | Y_j^{a,b},C, X_j \Big]  + \mu(a,b, C, X_j ; \gamma^* ) \Bigg] \\
 = & \ E \Bigg[ \frac{  Y_j^{a,b} - \mu(a,b,C,X_j  ; \gamma^*) }{ \pi(C,X_j ; \alpha_0 ) \psi(C,X_j ; \theta_0) }  E \Big[ \mathds{1}(A_j=a,A_{3-j}=b) | C, X_j \Big]  + \mu(a,b, C, X_j ; \gamma^* ) \Bigg]  & \mbox{ by \textbf{B1}} \\
    = & \ E \Bigg[ \frac{  Y_j^{a,b} -\mu(a,b,  C,X_j ; \gamma^* )}{ \pi(C,X_j ; \alpha_0 ) \psi(C,X_j ; \theta_0) }  P(A_j=a | C, X_j) P(A_{3-j}=b | C,X_j) + \mu(a,b, C, X_j ; \gamma^* ) \Bigg]  & \mbox{ by \textbf{B7}}   \\ 
  = & \ E \Big[  Y_j^{a,b} -\mu(a,b,  C,X_j ; \gamma^*  )   + \mu(a,b, C, X_j ; \gamma^* ) \Big] \\
   = & \ E \big[ Y_j^{a,b} \big] .
 \end{align*}

%%%%%%%%%%%%%%%%%%%%%%%%%%%%%%%%

\section{Data analysis and Simulations}

Here we describe the models that were used in the data analysis.  The Model 1 estimator uses predicted values from an outcome regression model, and from a joint propensity score model.  We use generalized additive models to model $E[ Y_j | A_j,A_{3-j}, C,X_j,X_{3-j}]$, the conditional mean of the twin's Drinking Index outcome, with the following model formulation:

\noindent  \texttt{Drinking.index $\sim$  s(Parent.alcohol.abuse) + s(Parent.drug.abuse) + } 

\texttt{ ti(Parent.alcohol.abuse, Parent.drug.abuse)  + Parent.occupation.level + }

\texttt{ Sex + Zygosity + Academic.motivation + Sex*Academic.motivation +   }

\texttt{ s(Conflict.with.parents)+ Age + Exposure + Parent.alcohol.abuse*Exposure +  }

\texttt{ Sex*Exposure + Cotwins.exposure +Zygosity*Cotwins.exposure + Exposure*Cotwins.exposure   }

\noindent where \texttt{s(  )} indicates a smooth function of the variable, and \texttt{ti(  ,  )} and indicates an interaction term consisting of a smooth function of the two variables.  In order to enforce symmetry between the Twin 1's and the Twin 2's, we fit this model on a stacked dataset combining the data for the Twin 1's and the data for the Twin 2's, so that the same fitted values of the regression parameters are used for both the Twin 1's and the Twin 2's.  

For the joint propensity score model, we first use generalized additive models to model logit $P(A_j=1 | C, X_j)$:

\medskip

\noindent \texttt{Exposure $\sim$ Parent.alcohol.abuse + Parent.drug.abuse + }

\texttt{ ti(Parent.alcohol.abuse, Parent.drug.abuse) + Externalizing.disorder + }

\texttt{ Academic.motivation + Conflict.with.parents + }

\texttt{ ti(Parent.alcohol.abuse, Conflict.with.parents) + s(Age), family=binomial }

\medskip

\noindent As before, we fit this propensity score model on a stacked dataset to enforce symmetry between the Twin 1's and the Twin 2's.  We then obtain predicted values for the joint distribution $F(a,b) :=P(A_1=a,A_2=b  | C,X_1,X_2 )$ from predicted values for the margins using a Dale model \cite{Dale}.  Under the Dale model, the joint distribution is determined by the two margins $F_1(0) :=P(A_1=0 | C,X_1)$ and $F_2(0):=P(A_2=0 | C,X_2)$ together with a parameter for the association, the cross ratio $\ds{\psi= \frac{ F(0,0) \times F(1,1) }{F(0,1) \times F(1,0)}  \in (0,\infty)}$, where $\psi=1$ corresponds to the case that $A_1$ and $A_2$ are conditionally independent.  Here we allow the strength of the association between $A_1$ and $A_2$ to depend on zygosity and sex, and we assume $\log( \psi_i) = \alpha_0 + \alpha_1 Zygosity_i + \alpha_2 Sex_i $.  We estimate $\alpha=(\alpha_0,\alpha_1,\alpha_2)$ using maximum likelihood, treating the margins as fixed.  Then $F(0,0)$ is determined by the margins $F_1(0)$ and $F_2(0)$ and the estimated $\widehat{\alpha}$:
\begin{align*}
 F ( &0,0) =  \\
 & \frac{1}{2( \psi-1)}  \Bigg\{  \Big[ 1+(F_1(0)+F_2(0))(\psi-1) \Big] + \sqrt{  \Big[ 1+(F_1(0)+F_2(0))(\psi-1) \Big]^2 + 4 \psi (1- \psi) F_1(0) F_2(0) }  \Bigg\}
 \end{align*}
if $\psi \neq 1$, and $F(0,0)= F_1(0)F_2(0)$ otherwise.   

\medskip

The Model 2 estimator uses predicted values from three models.  We use the same conditional mean model $E [ Y_j | A_j,A_{3-j}, C,X_j]$ and the same propensity score model $P(A_j | C,X_j)$ as above.  Finally, we model logit$P(A_j=1 | C,X_{3-j})$, where $P(A_j=1 | C,X_{3-j})$ is the probability that the twin is exposed given shared covariates and their co-twin's individual covariates, as:

\noindent \texttt{Exposure $\sim$ Parent.alcohol.abuse + Parent.drug.abuse + }

\texttt{ ti(Parent.alcohol.abuse, Parent.drug.abuse) +  }

\texttt{ Cotwins.externalizing.disorder + Cotwins.academic.motivation +  }

\texttt{ Cotwins.conflict.with.parents + }

\texttt{ ti(Parent.alcohol.abuse, Cotwins.conflict.with.parents) + s(Age), family=binomial }

\noindent and fit this model on the stacked dataset.

\medskip

Next, we describe the data generating mechanisms used in our simulation study.  Under the first data-generating mechanism, Model 1 is correctly specified but Model 2 is not.  For each simulated dataset, we resample $n=500$ twin pairs from the Minnesota Twin Family Study (MTFS) data to generate shared and individual baseline covariates.  Let the $s$th twin pair in the simulated data have covariates $(C_s,X_{s1},X_{s2})$.  We then use the model for $P(A_1,A_2 | C,X_1,X_2)$ described above to generate exposures, such that twin pair $s$ will have exposures $A_{s1}=a,A_{s2}=b$ with probability $P(A_1=a,A_2=b | C=C_s,X_1=X_{s1}, X_2=X_{s2})$, the predicted values from the model fit on the MTFS data.  The Drinking Index outcome in the MTFS data is approximately normal, so we generate outcomes from a bivariate normal distribution where the mean corresponds to predicted values of the outcome regression model fit on the MTFS data, and where values of the variance-covariance matrix are chosen to approximate the variance and covariance seen in the MTFS data.  In particular, we take outcomes for Twin 1 and Twin 2 to be more strongly correlated among MZ twins than among DZ twins.  Specifically, for twin pair $s$, we draw 

\[ ( Y_{s1},Y_{s2} )  \ \sim \ MVN_2 \left( \left[ \begin{array}{c} \mu_1 \\ \mu_2 \end{array} \right], \ \left( \begin{array}{cc} 8 & \sigma \\ \sigma & 8 \end{array} \right) \right), \] 
where $\mu_j= E[ Y_j | A_j=A_{sj}, A_{3-j}=A_{s,3-j}, C=C_s,X_j=X_{sj}, X_{3-j}=X_{s,3-j} ]$ is the predicted value from the model fit on the MTFS data, and where $\sigma= 3.5$ if twin pair $s$ is MZ, and $\sigma=1$ if twin pair $s$ is DZ.   

For the second data-generating mechanism, under which Model 1 and Model 2 are both correctly specified, we modify the step in which we generate exposures, drawing $A_{s1} \sim Bern(p_{s1})$, where $p_{s1}= P(A_1=1 | C=C_s,X_1=X_{s1})$, and independently drawing $A_2  \sim Bern(p_{s2})$, where $p_{s2}= P(A_2=1 | C=C_s,X_2=X_{s2})$.

\end{document}